\documentclass{elsart}
\usepackage{graphicx} 
\usepackage{epsfig}   
\usepackage{dcolumn}  
\usepackage{color}

\graphicspath{{ps}}

\renewcommand{\arraystretch}{1.1}

\begin{document}
\begin{frontmatter}

\epsfysize2.cm
\hspace{-9.5cm}
\epsfbox{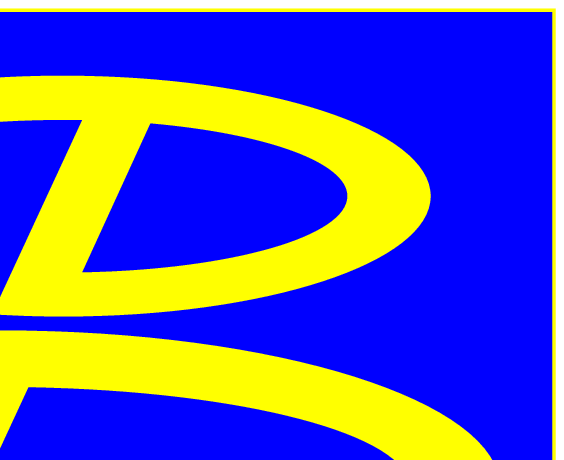}    
\begin{flushright}
\vskip -1.5cm
\noindent
\hspace*{9.0cm}Belle Prerpint 2006-10\\
\hspace*{9.0cm}KEK   Preprint 2006-4
\end{flushright}
\title{ \boldmath \quad\\[0.5cm] Measurements of branching fractions and $q^2$ distributions for 
$B \to \pi \ell \nu$ and $B \to \rho \ell \nu$ Decays with $B \to D^{(*)} \ell \nu$ Decay Tagging}


\collab{Belle Collaboration}
   \author[Nagoya]{T.~Hokuue}, 
   \author[KEK]{K.~Abe}, 
   \author[TohokuGakuin]{K.~Abe}, 
   \author[KEK]{I.~Adachi}, 
   \author[Tokyo]{H.~Aihara}, 
   \author[Tsukuba]{Y.~Asano}, 
   \author[ITEP]{T.~Aushev}, 
   \author[Sydney]{A.~M.~Bakich}, 
   \author[ITEP]{V.~Balagura}, 
   \author[Melbourne]{E.~Barberio}, 
% ...

% ...
\address[KEK]{High Energy Accelerator Research Organization (KEK), Tsukuba, Japan}
% ...
\address[ITEP]{Institute for Theoretical and Experimental Physics, Moscow, Russia}
% ...
\address[Melbourne]{University of Melbourne, Victoria, Australia}
\address[Nagoya]{Nagoya University, Nagoya, Japan}
% ...
\address[Sydney]{University of Sydney, Sydney, NSW, Australia}
% ...
\address[TohokuGakuin]{Tohoku Gakuin University, Tagajo, Japan}
% ...
\address[Tokyo]{Department of Physics, University of Tokyo, Tokyo, Japan}
% ...
\address[Tsukuba]{University of Tsukuba, Tsukuba, Japan}
% ...

\vspace{1.0cm}

\begin{abstract}
We report measurements of the charmless semileptonic decays 
$B^0 \to \pi^- / \rho^- \ell^{+} \nu$ and $B^+ \to \pi^0 / \rho^0 \ell^{+} \nu$,  
based on a sample of $2.75 \times 10^8$ $B \bar{B}$ events collected at the $\Upsilon(4S)$ resonance 
with the Belle detector at the KEKB $e^+e^-$ asymmetric collider.
In this analysis, the accompanying $B$ meson is reconstructed in the
semileptonic mode $B \rightarrow D^{(*)} \ell \nu$, enabling detection of the signal modes with high 
purity.
We measure the branching fractions
${\mathcal B}(B^0 \to \pi^- \ell^+ \nu)  = 
(1.38\pm 0.19\pm 0.14\pm 0.03) \times 10^{-4}$,
${\mathcal B}(B^0 \to \rho^- \ell^+ \nu)  = 
(2.17\pm 0.54\pm 0.31\pm 0.08) \times 10^{-4}$,
${\mathcal B}(B^+ \to \pi^0 \ell^+ \nu)  = 
(0.77\pm 0.14\pm 0.08\pm 0.00) \times 10^{-4}$ and
${\mathcal B}(B^+ \to \rho^0 \ell^+ \nu)  = 
(1.33\pm 0.23\pm 0.17\pm 0.05) \times 10^{-4}$,
where the errors are statistical, experimental systematic, and systematic due
to form-factor uncertainties, respectively.
For each mode we also present the partial branching fractions in three $q^2$ intervals: 
$q^2 < 8$, $8 \leq q^2 < 16$, and $q^2 \geq 16$\,GeV$^2/c^2$.
From our partial branching fractions for $B \to \pi \ell \nu$ 
and recent results for the form factor from unquenched Lattice QCD calculations, 
we obtain values of the CKM matrix element $|V_{ub}|$.
\end{abstract}

\end{frontmatter}


{\renewcommand{\thefootnote}{\fnsymbol{footnote}}}
\setcounter{footnote}{0}

\section{Introduction}
\label{sec:Introduction}
Exclusive $B \to X_u \ell \nu$ decays proceed dominantly via a $b \to u W^-$ tree 
process 
and can be used to determine $|V_{ub}|$, 
one of the smallest and least known elements of the 
Cabibbo-Kobayashi-Maskawa matrix~\cite{KM}.
However, the need to translate the observed rate to a $|V_{ub}|$ value 
using model-dependent decay form-factors (FF) has resulted in large theoretical uncertainties.
The recent release of FF results for $B \to \pi \ell \nu$ calculated by unquenched Lattice QCD (LQCD)~\cite{FNAL04,HPQCD04} 
makes possible the first model-independent determination of $|V_{ub}|$.
Since LQCD results are available only in the high $q^2$ region ($\geq 16$\,GeV$^2/c^2$), 
a clean measurement of the partial $B \to \pi \ell \nu$ branching fraction in the same high $q^2$ region is needed.

There have been several measurements in the past by CLEO, BaBar and
Belle for the $B \to \pi \ell \nu$, $\rho \ell \nu$, $\eta \ell \nu$
and $\omega \ell \nu$ modes~\cite{CLEO1996,CLEO2000,CLEO2003,BABAR2003,BELLE2004,BABAR2005}.
The analyses in these measurements utilize the method, originally developed by CLEO, where the 
$B$ decays are reconstructed by inferring the undetected neutrino 
mass from missing energy and momentum (``$\nu$-reconstruction 
method'')~\cite{CLEO1996}.
In the $B$-factory era, we will improve the statistical 
precision by simply applying the $\nu$-reconstruction method using a large amount of data.
However, the poor signal-to-noise ratio will limit the systematic uncertainty of the measurement.

In this paper we present measurements of $B^0 \to \pi^- / \rho^- \ell^+ \nu$ 
and $B^+ \to \pi^0 / \rho^0 \ell^+ \nu$ decays using $B \to D^{(*)} \ell \nu$ decay tagging.
We reconstruct the entire decay chain from the $\Upsilon(4S)$,
$\Upsilon(4S) \to B_{\rm sig}B_{\rm tag}$, $B_{\rm sig} \to \pi / \rho \ell \nu$ and
$B_{\rm tag} \to D^{(*)} \ell \bar{\nu}$ with several $D^{(*)}$ sub-modes.
The back-to-back correlation of the two $B$ mesons in the $\Upsilon(4S)$
rest frame allows us to constrain the kinematics of the double semileptonic decay. 
The signal is reconstructed in four modes, $B^0 \to \pi^- / \rho^- \ell^+ \nu$ and $B^+ \to \pi^0 / \rho^0 \ell^+ \nu$.
Yields and branching fractions are extracted from a simultaneous fit of the $B^0$ and $B^+$ samples 
in three intervals of $q^2$, accounting for cross-feed between modes as well as other backgrounds.
We have applied this method to $B \to \pi / \rho \ell \nu$ decays
for the first time, and have succeeded in reconstructing these decays 
with significantly improved signal-to-noise ratios compared to the $\nu$-reconstruction method.
Inclusion of charge conjugate decays is implied throughout this paper.

\section{Data Set and Experiment}
\label{sec:data_exp}

The analysis is based on data recorded with the Belle detector at the KEKB collider 
operating at the center-of-mass (c.m.) energy for the $\Upsilon(4S)$ resonance~\cite{KEKB}.
The $\Upsilon(4S)$ dataset that is used corresponds to an
integrated luminosity of 253 fb$^{-1}$ and contains $2.75 \times 10^8$ 
$B \bar{B}$ events.

The Belle detector is a large-solid-angle magnetic spectrometer
that consists of a silicon vertex detector (SVD),
a 50-layer central drift chamber (CDC), 
an array of aerogel threshold \v{C}erenkov counters (ACC),
a barrel-like arrangement of time-of-flight scintillation counters (TOF),
and an electromagnetic calorimeter comprised of CsI(Tl) crystals (ECL)
located inside a super-conducting solenoid coil 
that provides a 1.5~T magnetic field.  
An iron flux-return located outside of the coil is instrumented
to detect $K_L^0$ mesons and to identify muons (KLM).  
The detector is described in detail elsewhere~\cite{BELLE}. 
Two inner detector configurations were used. A 2.0 cm beam pipe
and a 3-layer silicon vertex detector was used for the first sample
of $152 \times 10^6$ $B\bar{B}$ pairs, while a 1.5 cm beam pipe, a 4-layer
silicon detector, and a small-cell inner drift chamber were used to record  
the remaining 123 million $B\bar{B}$ pairs~\cite{Ushiroda}.  

A detailed Monte Carlo (MC) simulation, which fully describes the detector
geometry and response and is based on GEANT~\cite{GEANT}, is
applied to estimate the signal detection efficiency and to study the 
background.
To examine the FF dependence, MC samples for the 
$B \to \pi / \rho \ell \nu$ signal decays are generated 
with different form-factor models:
a quark model (ISGW~II ~\cite{ISGW2}), 
light cone sum rules (LCSR for $B \to \pi \ell \nu$ ~\cite{Ball04_pi} 
and $B \to \rho \ell \nu$ ~\cite{Ball04_rho}) and 
quenched lattice QCD (UKQCD ~\cite{UKQCD}). We also use unquenched lattice QCD
(FNAL~\cite{FNAL04} and HPQCD~\cite{HPQCD04}) for $B \to \pi \ell \nu$ and 
a relativistic quark model (Melikhov ~\cite{Melikhov}) for $B \to \rho \ell \nu$.
To model the cross-feed from other $B \to X_u \ell \nu$ decays,
MC samples are generated with the ISGW~II model for the resonant components 
($B \to \pi \ell \nu$ and $B \to \rho \ell \nu$ components are excluded in this sample) 
and the DeFazio-Neubert model ~\cite{Fazio-Neubert} for 
the non-resonant component.
To model the $B\bar{B}$ and continuum backgrounds, large generic 
$B\bar{B}$ and $q\bar{q}$ Monte Carlo (based on Evtgen~\cite{Evtgen}) samples are used.

\section{Event Reconstruction and Selection}

Charged particle tracks are reconstructed from hits in the SVD and the CDC. 
They are required to satisfy track quality cuts based on their impact 
parameters relative to the measured interaction point (IP) of the two beams. 
Charged kaons are identified by combining information on ionization loss 
($dE/dx$) in the CDC, \v{C}herenkov light yields in the ACC and time-of-flight 
measured by the TOF system.
For the nominal requirement, the kaon identification efficiency is 
approximately $88\%$ and the rate for misidentification of pions as 
kaons is about $8\%$.
Hadron tracks that are not identified as kaons are treated as pions.
Tracks satisfying the lepton identification criteria, as described below, 
are removed from consideration.

Neutral pions are reconstructed using $\gamma$ pairs with an invariant mass 
between 117 and 150\,MeV/$c^2$.
Each $\gamma$ is required to have a minimum energy of $50$~MeV.
$K_S^0$ mesons are reconstructed using pairs of tracks that are
consistent with having a common vertex and that have an invariant mass 
within $\pm 12$\,MeV/$c^2$ of the known $K_S^0$ mass.

Electron identification is based on a combination of $dE/dx$ in the CDC,
the response of the ACC, shower shape in the ECL and the ratio of energy 
deposit in the ECL to the momentum measured by the tracking system.
Muons are identified by their signals in the KLM resistive plate counters, which are interleaved 
with the iron of the solenoid return yoke.
The lepton identification efficiencies are estimated to be about 90\% 
for both electrons and muons in the momentum region above 1.2\,GeV/$c$, 
where leptons from prompt $B$ decays dominate.
The hadron misidentification rate is measured 
using reconstructed $K_S^0 \to \pi^+ \pi^-$
and found to be less than 0.2\% for electrons and 1.5\% for muons
in the same momentum region.

For the reconstruction of $B_{\rm tag} \to D^{(*)} \ell \bar{\nu}$, 
the lepton candidate is required to have the correct sign charge with
respect to the $D$ meson flavor and a laboratory momentum ($p_{\ell}^{lab}$) greater
than 1.0\,GeV/$c$.
The $D$ meson candidates are reconstructed by using seven decay modes of $D^+$:
$D^+ \to K^- \pi^+ \pi^+$, $K^- \pi^+ \pi^+ \pi^0$, $K_S^0 \pi^+$, 
$K_S^0 \pi^+ \pi^0$, $K_S^0 \pi^+ \pi^+ \pi^-$,
$K^+ K_S^0$, $K^+ K^- \pi^+$; and ten decay modes of $D^0$: 
$D^0 \to K^- \pi^+$, $K^- \pi^+ \pi^0$, $K^- \pi^+ \pi^+ \pi^-$, $K_S^0 \pi^0$, 
$K_S^0 \pi^+ \pi^-$, $K_S^0 \pi^+ \pi^- \pi^0$, $K^- \pi^+ \pi^+ \pi^- \pi^0$,
$K^+ K^-$, $K_S^0 K^+ K^- $, $K_S^0 K^- \pi^+$.
The candidates are required to have an invariant mass $m_D$ within
$\pm 2.5\sigma$ ($\sigma$ is a standard deviation) of the nominal $D$ mass, 
where the mass resolution $\sigma$ is dependent on the decay mode. 
$D^{*}$ mesons are reconstructed in the modes $D^{*+} \to D^0 \pi^+$, $D^+ \pi^0$ 
and $D^{*0} \to D^0 \pi^0$ by combining a $D$ meson candidate and 
a charged or neutral pion.
Each $D^*$ candidate is required to have a mass difference 
$\Delta m = m_{\bar{D}\pi} - m_{\bar{D}}$ 
within $\pm 2.5 \sigma$ of the nominal values.    

For the reconstruction of $B_{\rm sig} \to X_u \ell \nu$, 
the lepton candidate is required to have the right sign charge with
respect to the $X_u$ system and $p_{\ell}^{lab}$ greater than 0.8 GeV/$c$.
The $X_u$ system may consist of one pion or two pions (
$N_{\pi^+}=1$ or $N_{\pi^+}=N_{\pi^0}=1$ for a $\bar{B^0}$ tag and 
$N_{\pi^0}=1$ or $N_{\pi^+}=N_{\pi^-}=1$ for a $B^-$ tag).
The event is required to have no additional charged tracks or $\pi^0$ 
candidates.
We also require that the residual energy from neutral clusters be
less than 0.15 GeV ($E_{\rm neut} < 0.15$\,GeV).   
The two leptons on the tag and the signal sides are required to have 
opposite charge.
The loss of signal due to $B^0 - \bar{B^0}$ mixing is estimated by 
MC simulation.

\begin{figure}[htbp]
 \begin{center}
  \mbox{\psfig{figure=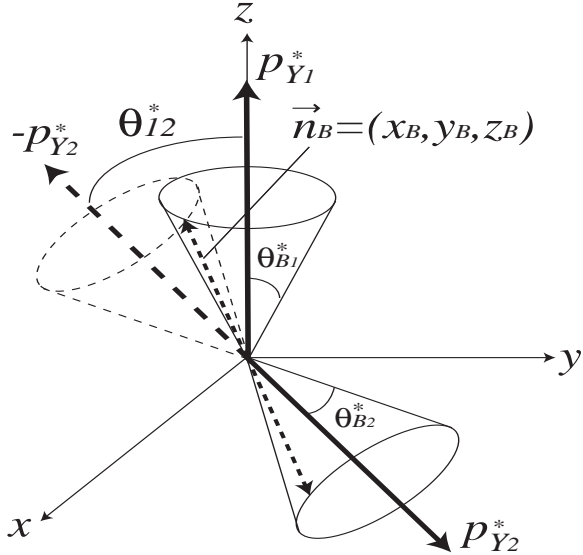,width=2.5in, height=2.5in, angle=0, scale=1.2 } } 
  \caption{Kinematics of the double semileptonic decay.}
  \label{fig:double_cone}
  \end{center}
\end{figure}

We then impose a constraint based on the kinematics of the double semileptonic decay
in the $\Upsilon(4S)$ rest frame.
In the semileptonic decay on each side, $B_{1(2)} \to Y_{1(2)} \nu$ 
($Y_1 = D^{(*)} \ell$ and $Y_2 = X_u \ell$), the angle between the 
$B_{1(2)}$ meson and the detected $Y_{1(2)}$ system 
$\theta_{B_{1(2)}}$ is calculated from the relation,
${P_{\nu}^{*}}^2 = (P_{B}^{*} - P_{Y}^{*})^2 = 0$ ($P^*$: 4-momentum vector)
and the known $p_B^*$ (the absolute momentum of the mother $B$ meson).
This means that the $B_{1(2)}$ direction is constrained on the surface of a 
cone defined with the angle $\theta_{B_{1(2)}}^*$ around the direction 
of the $Y_{1(2)}$ system, as shown graphically in 
Fig.~\ref{fig:double_cone}. The back-to-back relation of the two $B$ meson directions then implies that
the real $B$ direction is on the intersection of the two cones when one of
the $B$ systems is spatially inverted.
Denoting $\theta_{12}^*$ the angle between the $p_{Y1}^*$ and $-p_{Y2}^*$,
the $B$ directional vector $\vec{n}_B = (x_B, y_B, z_B)$ 
is given by, 
$z_B = \mbox{cos}\theta_{B_1}^*$,  
$y_B = (\mbox{cos}\theta_{B_2}^* - \mbox{cos}\theta_{B_2}^*\mbox{cos}\theta_{12}^*)
     / \mbox{sin}\theta_{12}^*$,
and
\begin{eqnarray}
 {x_B}^2 = 1-\frac{1}{\mbox{sin}^2\theta_{12}^*}
(\mbox{cos}^2\theta_{B_1}^* + \mbox{cos}^2\theta_{B_2}^* 
- 2 \mbox{cos}\theta_{B_1}^*\mbox{cos}\theta_{B_2}^*\mbox{cos}\theta_{12}^*)
 \label{eq:x_B}
\end{eqnarray}
with the coordinate definition in Fig.~\ref{fig:double_cone}, where the $p_{Y1}^*$ and $p_{Y2}^*$ are aligned along 
the $z$-axis and in the $y-z$ plane, respectively.
If the hypothesis of the double semileptonic decay is correct and all
the decay products are detected except for the two neutrinos, $x_B^2$
must range from 0 to 1.
Events passing a rather loose cut $x_B^2 > -2.0$ are used for signal 
extraction at a later stage of the analysis.

Since the direction of the $B$ meson is not uniquely determined,
we calculate, $q^2$ as $q^{2} = (E^{*}_{\rm beam} - E^{*}_{X_u})^2 - p^{*}_{X_u}{^2}$, 
using the beam energy ($E^*_{\rm beam}$), energy ($E^*_{X_u}$) and 
momentum ($p^*_{X_u}$) of the $X_u$ system and neglecting the momentum of the 
$B$ meson in the c.m. system.
The signal Monte Carlo simulation finds that the $q^2$ resolution
depends on the reconstructed $q^2$ and is in the range 0.32-0.95\,GeV$^2/c^2$.

According to Monte Carlo simulation, the largest backgrounds originate from 
$B \to X_c \ell \nu$ and non-signal $B \to X_u \ell \nu$ decays, 
where some particles escape detection.
There are sizable contributions from cross talk between the $\bar{B^0}$ and
$B^+$ tags.
The contribution from $q\bar{q}$ processes is found to be negligible.

For events selected as described above, the signal MC simulation 
indicates that the total detection efficiency ($\epsilon_{\rm total}$),
 averaged over electron and muon channels, is
$1.98\times 10^{-3}$ for $B^0 \to \pi^- \ell^+ \nu$ and
$0.76\times 10^{-3}$ for $B^0 \to \rho^- \ell^+ \nu$,
$1.49\times 10^{-3}$ for $B^+ \to \pi^0 \ell^+ \nu$ and
$1.78\times 10^{-3}$ for $B^+ \to \rho^0 \ell^+ \nu$
assuming the LCSR FF model.
Here, $\epsilon_{\rm total}$ is defined with respect to the number of $B \bar{B}$ pairs, 
where one $B$ decays into the signal mode, 
and includes the loss of signal due to $B^0 - \bar{B^0}$ mixing.
Because of the loose lepton momentum cut~($p_{\ell}^{lab}$ $>0.8$\,GeV/$c$), the variation 
of efficiency with different FF models is relatively small.
Table~\ref{tbl:Matrix_sig} gives the matrix $\epsilon(q^2_{\rm rec.}, q^2_{\rm true})$, the efficiency for a signal event 
generated with true $q^2$ in the bin $q^2_{\rm true}$ to be reconstructed in the bin $q^2_{\rm rec.}$.


\begin{table}[htbp]
 \begin{center}
  \caption{ Detection efficiency matrix $\epsilon(q^2_{\rm rec.}, q^2_{\rm true})$ based on the LCSR model in units of $10^{-3}$.}
  \label{tbl:Matrix_sig}
  {\tabcolsep = 1pt \footnotesize
  \begin{tabular}{l|ccc|ccc|ccc|ccc}
    \hline\hline
               &    \multicolumn{12}{c}{$q^2_{\rm rec.}$(GeV$^2/c^2$)} \\
    \hline
    $q^2_{\rm true}$  &       & $ \pi^- \ell^+ \nu $ &           &       & $\rho^- \ell^+ \nu $ &           &       & $\pi^0 \ell^+ \nu $ &           &       & $\rho^0 \ell^+ \nu $ & \\
    \cline{2-13}
    (GeV$^2/c^2$) & $< 8$ & $ 8 - 16$            & $\geq 16$ & $< 8$ & $ 8 - 16$             & $\geq 16$ & $< 8$ & $ 8 - 16$            & $\geq 16$ & $< 8$ & $ 8 - 16$             & $\geq 16$ \\
     \hline
         $< 8$ &  1.71 & 0.21                 & 0.00      &  0.59 & 0.03                  & 0.00      &  1.27 & 0.07                 & 0.00      &  1.50 & 0.08                  & 0.00      \\  
      $8 - 16$ &  0.05 & 1.82                 & 0.24      &  0.07 & 0.65                  & 0.05      &  0.10 & 1.43                 & 0.06      &  0.10 & 1.71                  & 0.13      \\   
     $\geq 16$ &  0.00 & 0.03                 & 1.89      &  0.02 & 0.13                  & 0.81      &  0.01 & 0.09                 & 1.45      &  0.01 & 0.08                  & 1.82      \\   
    \hline\hline
  \end{tabular}}
 \end{center}
\end{table}

\begin{figure}[htbp]
\vspace{1.0cm}
  \begin{center}
   \begin{tabular}{cc}
    \hspace{-0.0cm}{\mbox{\psfig{figure=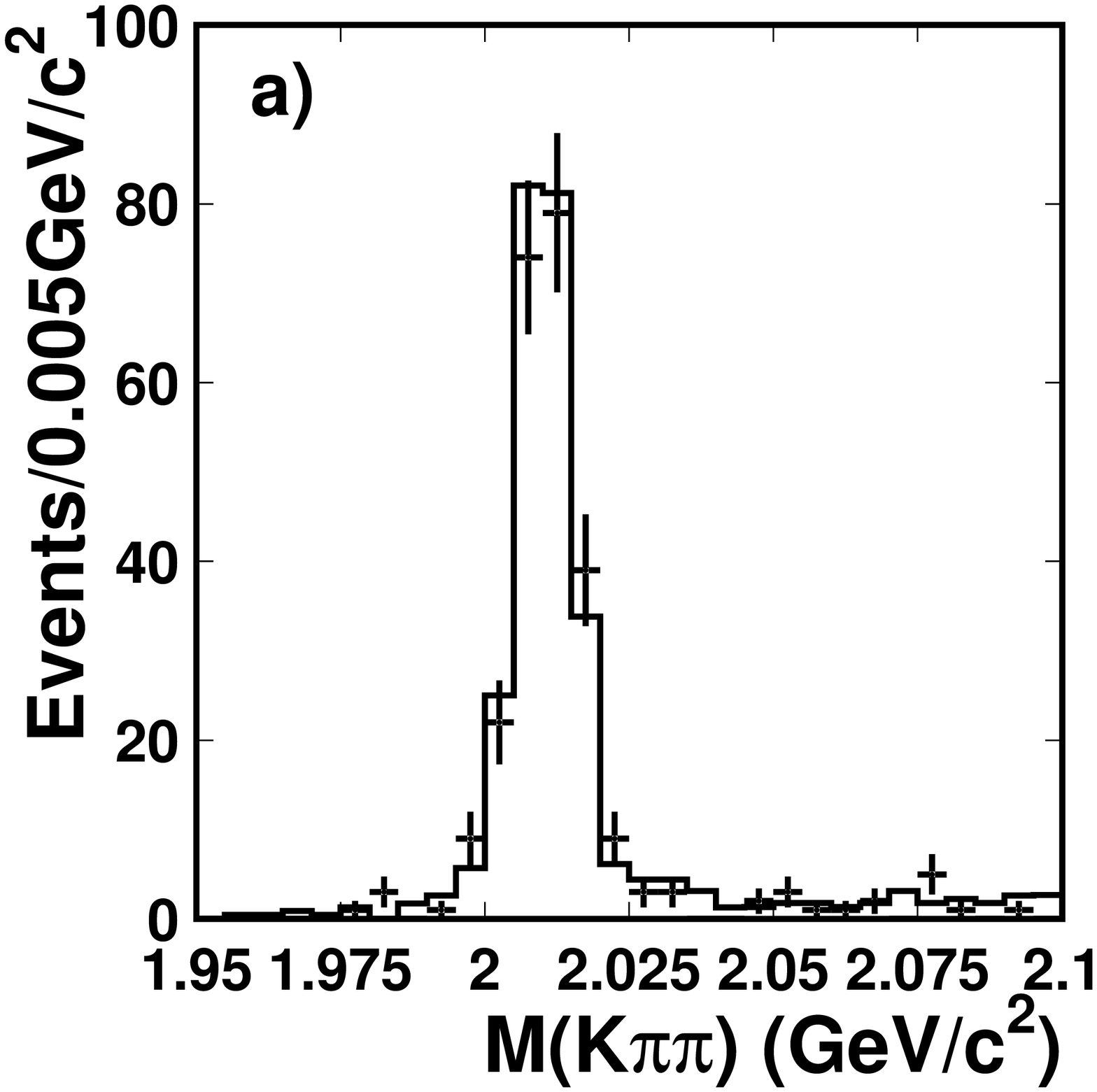,width=2.5in, height=2.5in, angle=0, scale=0.9 }}} &
    \hspace{-0.0cm}{\mbox{\psfig{figure=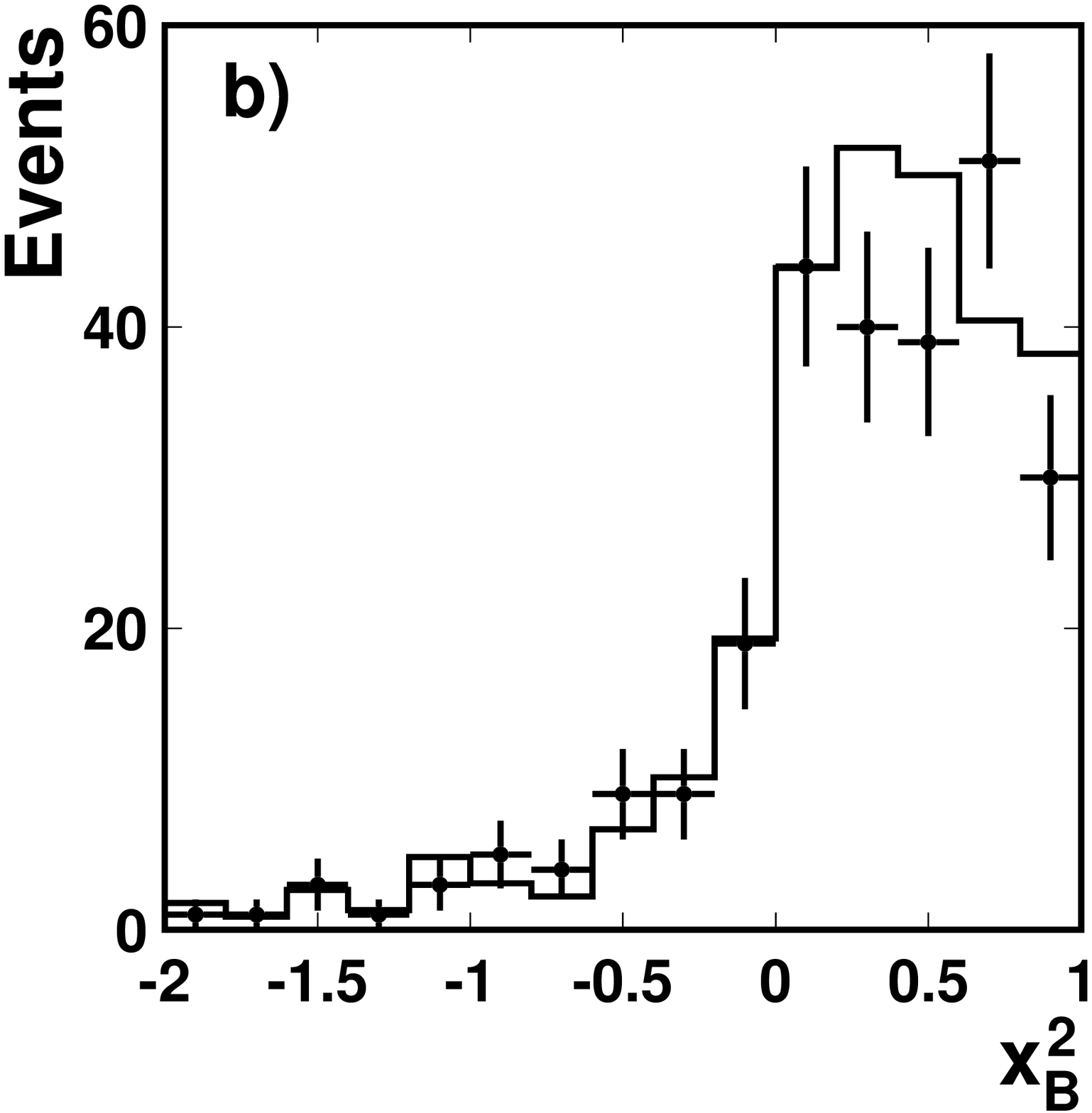,width=2.5in, height=2.5in, angle=0, scale=0.9 }}} \\
    \hspace{-0.0cm}{\mbox{\psfig{figure=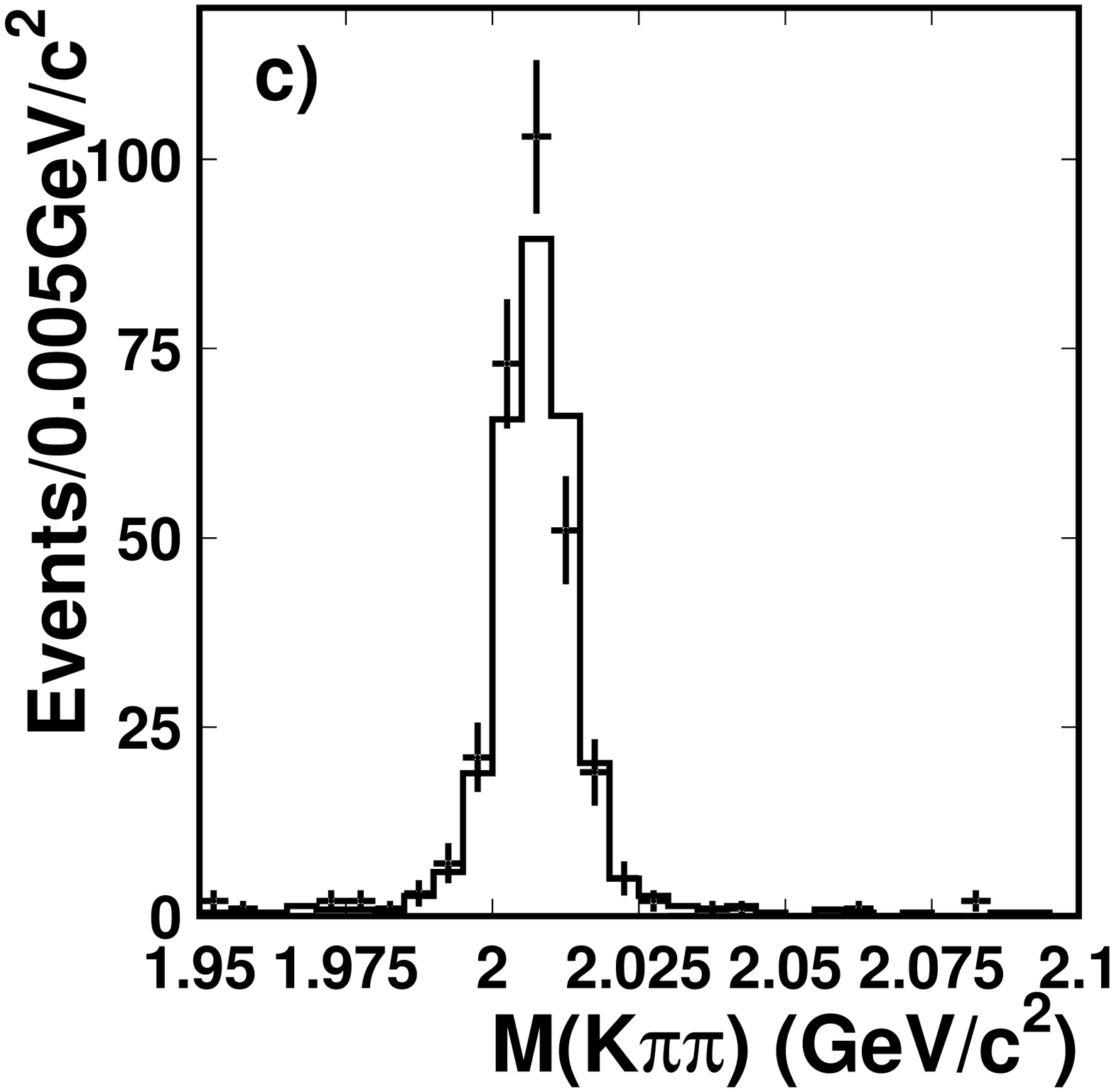,width=2.5in, height=2.5in, angle=0, scale=0.9 }}} &
    \hspace{-0.0cm}{\mbox{\psfig{figure=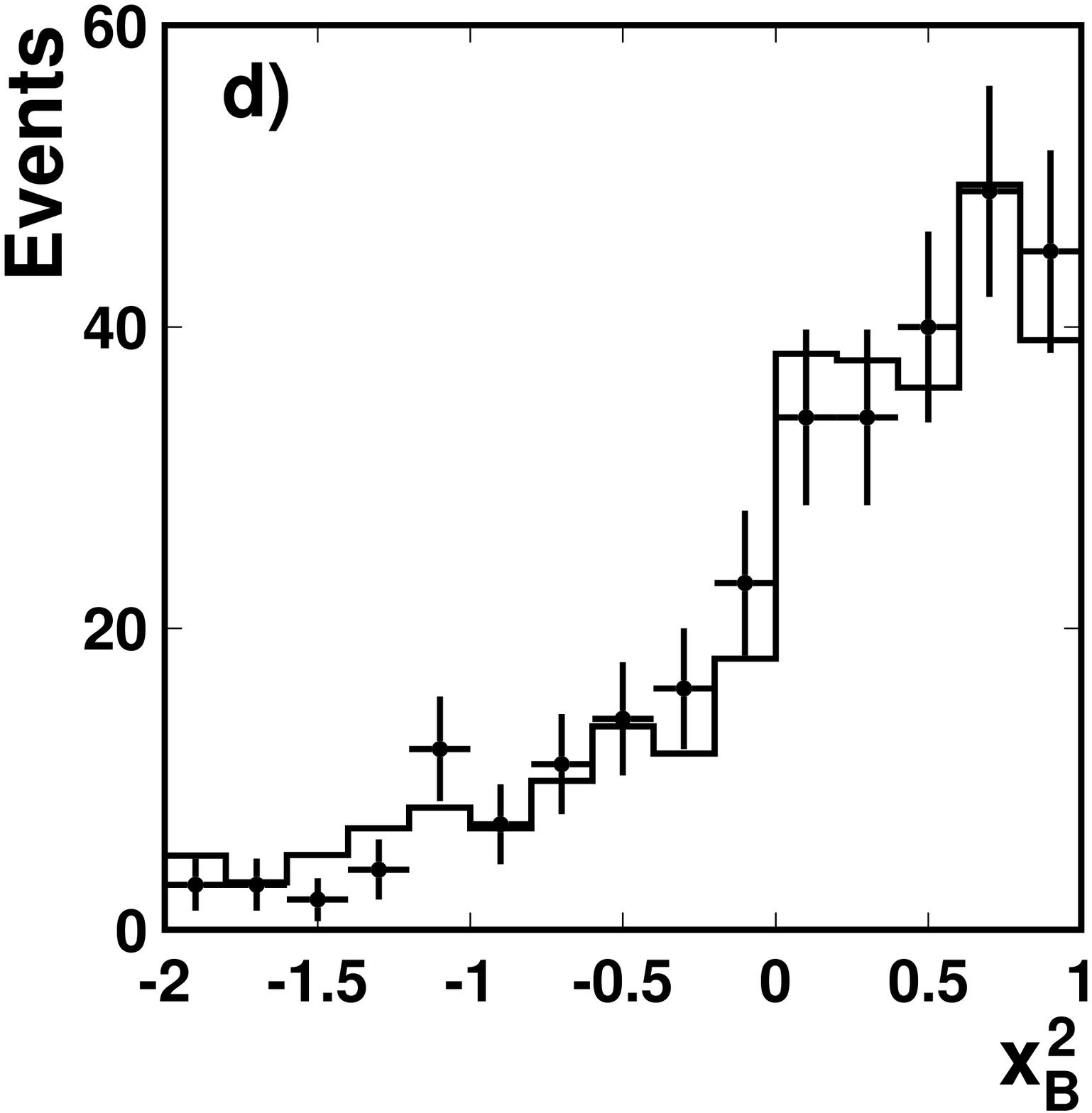,width=2.5in, height=2.5in, angle=0, scale=0.9 }}} \\
   \end{tabular}
   \caption{ Reconstructed $M(K\pi\pi)$ distribution (a) and $x^2_B$ distribution (b) 
   for the $B^0 \to D^{*-} \ell^{+} \nu$ calibration decay, (c) and (d) are for the $B^+ \to D^{*0} \ell^{+} \nu$ decay;
   points with error bars are data and the histogram is the signal MC.}
    \label{fig:dstlnu}
  \end{center}
\end{figure}

To check the validity of the method, we apply the procedure described above 
to reconstruct $B_{\rm sig}^0 \to D^{*-} \ell^{+} \nu$ followed 
by $D^{*-} \to \bar{D^0} \pi^{-}, \bar{D^0} \to K^{+} \pi^{-}$ for a $\bar{B}^0$ tag 
and $B_{\rm sig}^+ \to \bar{D}^{*0} \ell^{+} \nu$ followed by $\bar{D}^{*0} \to \bar{D^0} \pi^{0}, \bar{D^0} \to K^{+} \pi^{-}$ for a $B^-$ tag,
with the same requirement on the tagging side.
Figures~\ref{fig:dstlnu}-a) and c) show the $M_{K\pi\pi}$ distributions 
that are obtained in data and expected from MC.
As a result, we obtained $224.7 \pm 15.4$ ($295.9 \pm 17.6$) $\bar{B}^0 ~(B^-)$ tagged events.
These values are in good agreement with expected values $224.5 \pm 9.5$ ($288.6 \pm 11.7$) calculated from
the branching fractions ${\mathcal B}(B^{0} \to D^{*-} \ell^{+} \nu)$, ${\mathcal B}(D^{*-(0)} \to \bar{D^0} \pi^{-(0)})$ 
and ${\mathcal B}(\bar{D^0} \to K^{+} \pi^{-})$ in \cite{PDG2005} and efficiencies obtained from MC.
Here, we use ${\mathcal B}(B^+ \to \bar{D}^{*0} \ell^{+} \nu)$ calculated from ${\mathcal B}(B^{0} \to D^{*-} \ell^{+} \nu)$ 
and the liftime ratio~\cite{PDG2005}; ${\mathcal B}(B^+ \to \bar{D}^{*0} \ell^{+} \nu) = {\mathcal B}(B^{0} \to D^{*-} \ell^{+} \nu) \times (\tau_{B^+} / \tau_{B^0})$.
The ratio of the reconstructed to expected value, $R=1.00 \pm 0.08 \pm 0.05 ~(1.03 \pm 0.07 \pm 0.05)$ 
where the first error is statistical error and the second is due to the uncertainty of the branching fractions 
from \cite{PDG2005}, is consistent with unity.
Figures~\ref{fig:dstlnu}-b) and d) show a comparison of the reconstructed 
$x_B^2$ distribution in the above data samples with MC simulation. Data and MC are in good agreement.

\section{Extraction of Branching Fractions}
\label{sec:SignalExtraction}
\begin{figure}[htbp]
 \begin{center}
  \begin{tabular}{ccc}
   \hspace{-0.5cm}{\mbox{\psfig{figure=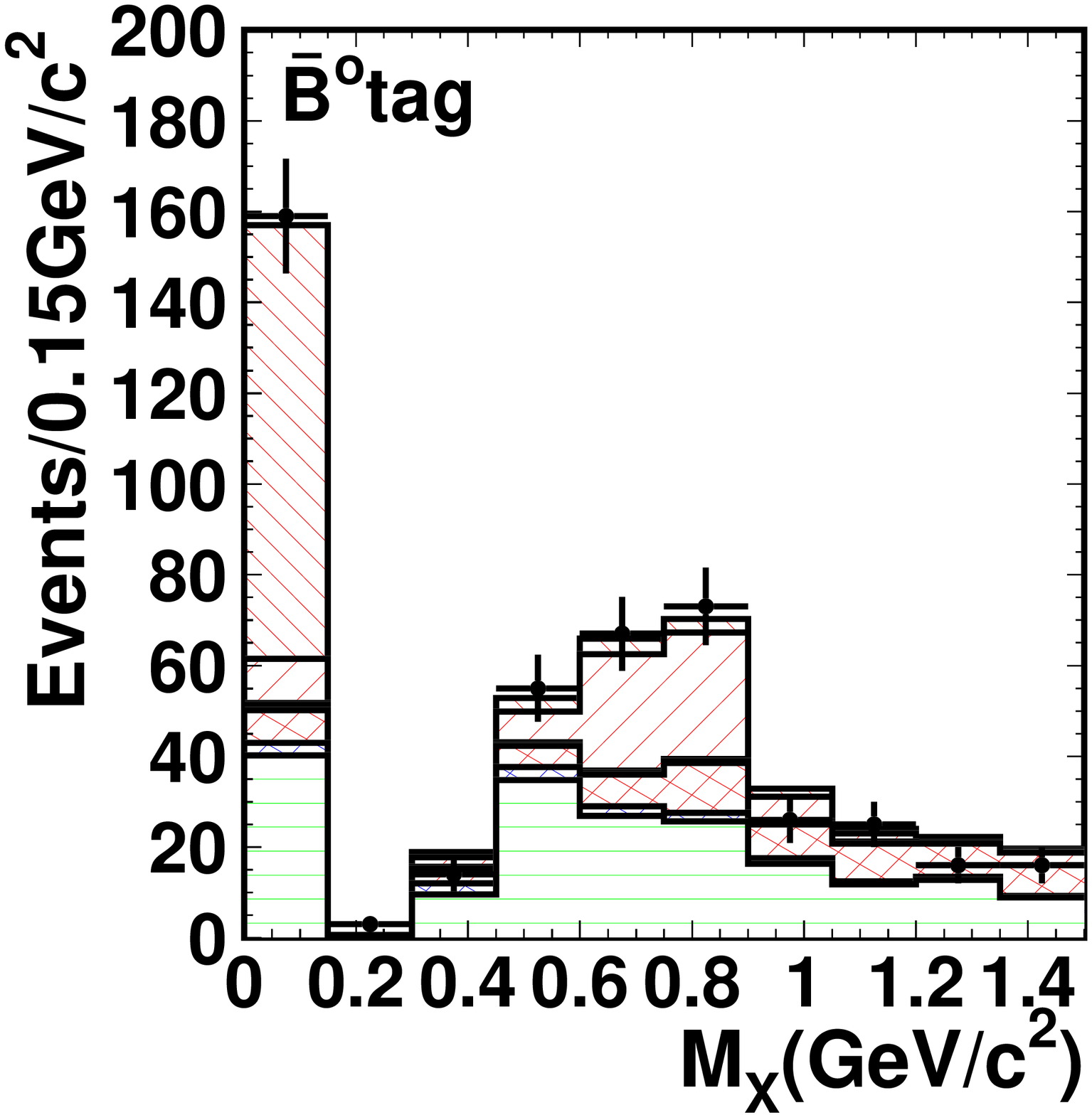,width=1.9in, height=2.0in, angle=0, scale=1.0 }} } &
   \hspace{-0.7cm}{\mbox{\psfig{figure=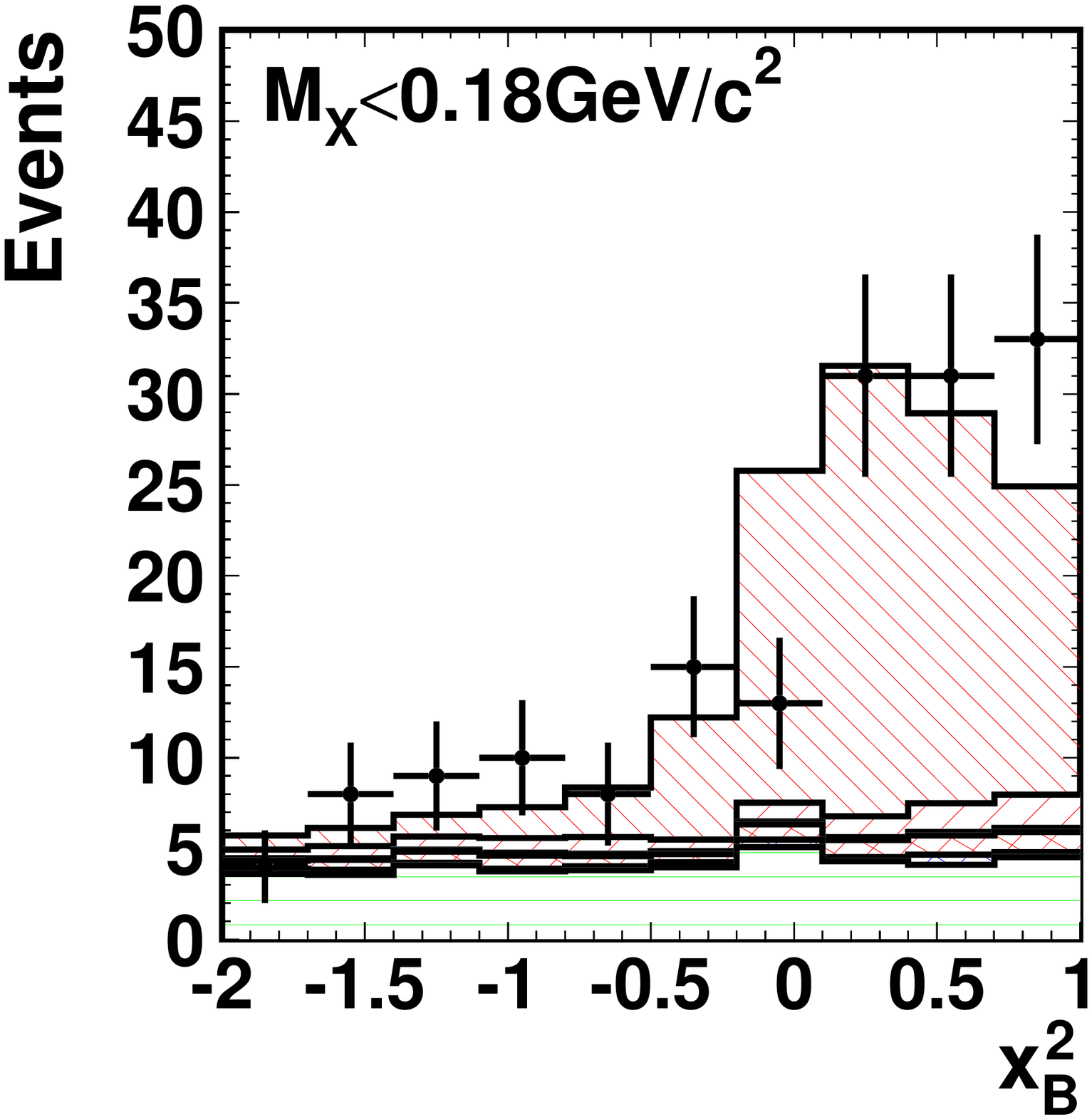,width=1.9in, height=2.0in, angle=0, scale=1.0 }} } &
   \hspace{-0.7cm}{\mbox{\psfig{figure=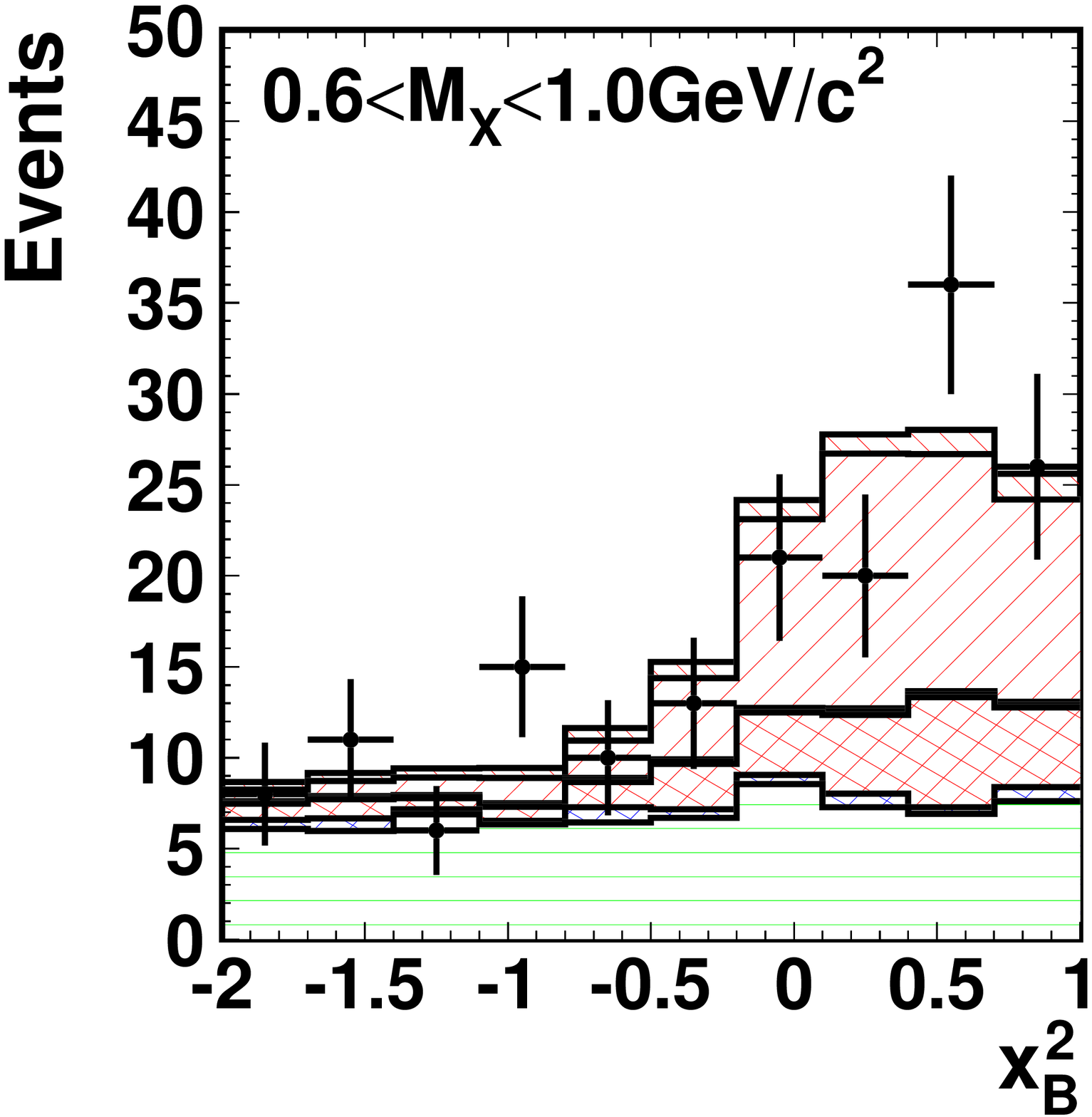,width=1.9in, height=2.0in, angle=0, scale=1.0}} } \\
   \hspace{-0.5cm}{\mbox{\psfig{figure=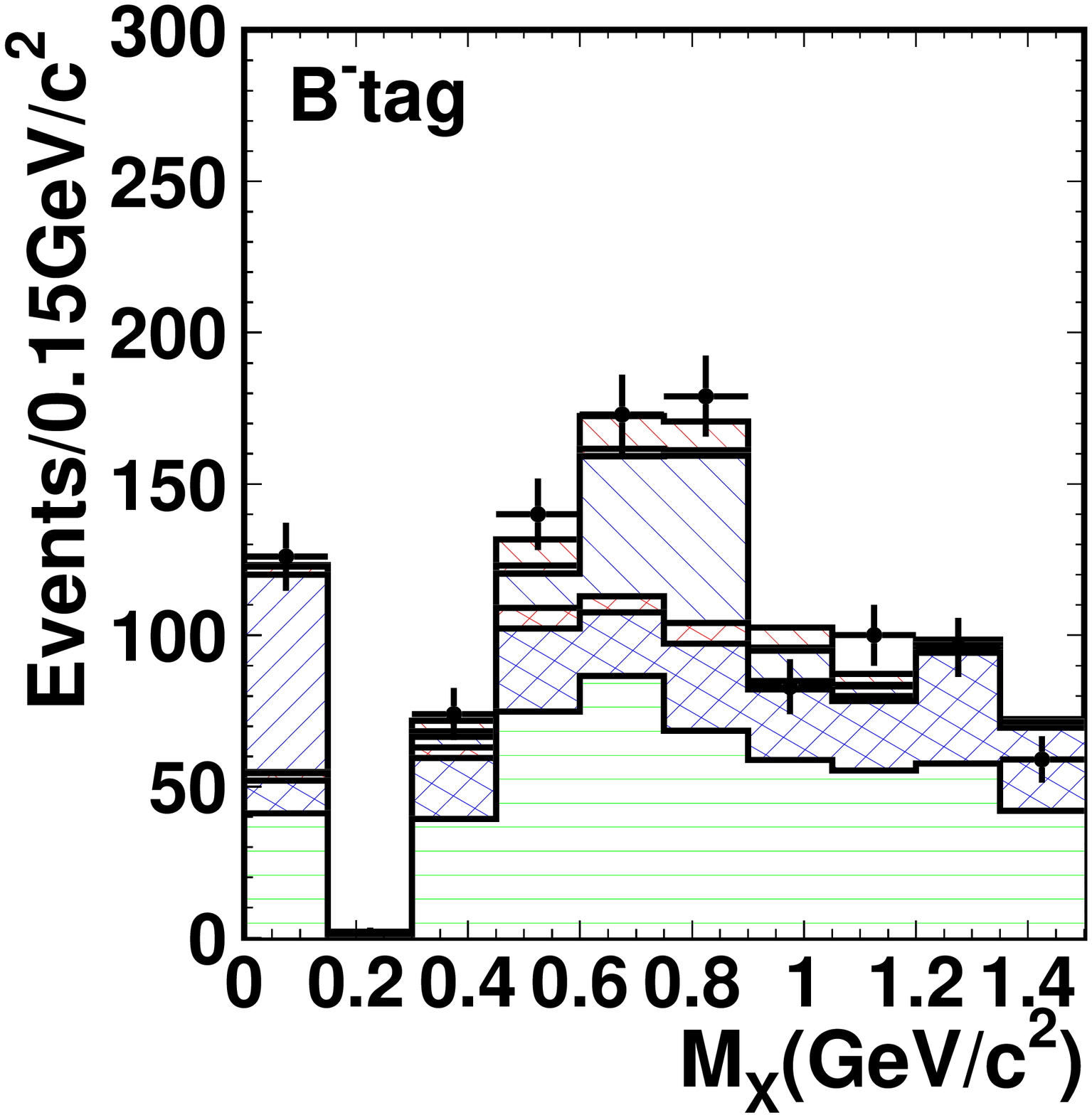,width=1.9in, height=2.0in, angle=0, scale=1.0 }} } &
   \hspace{-0.7cm}{\mbox{\psfig{figure=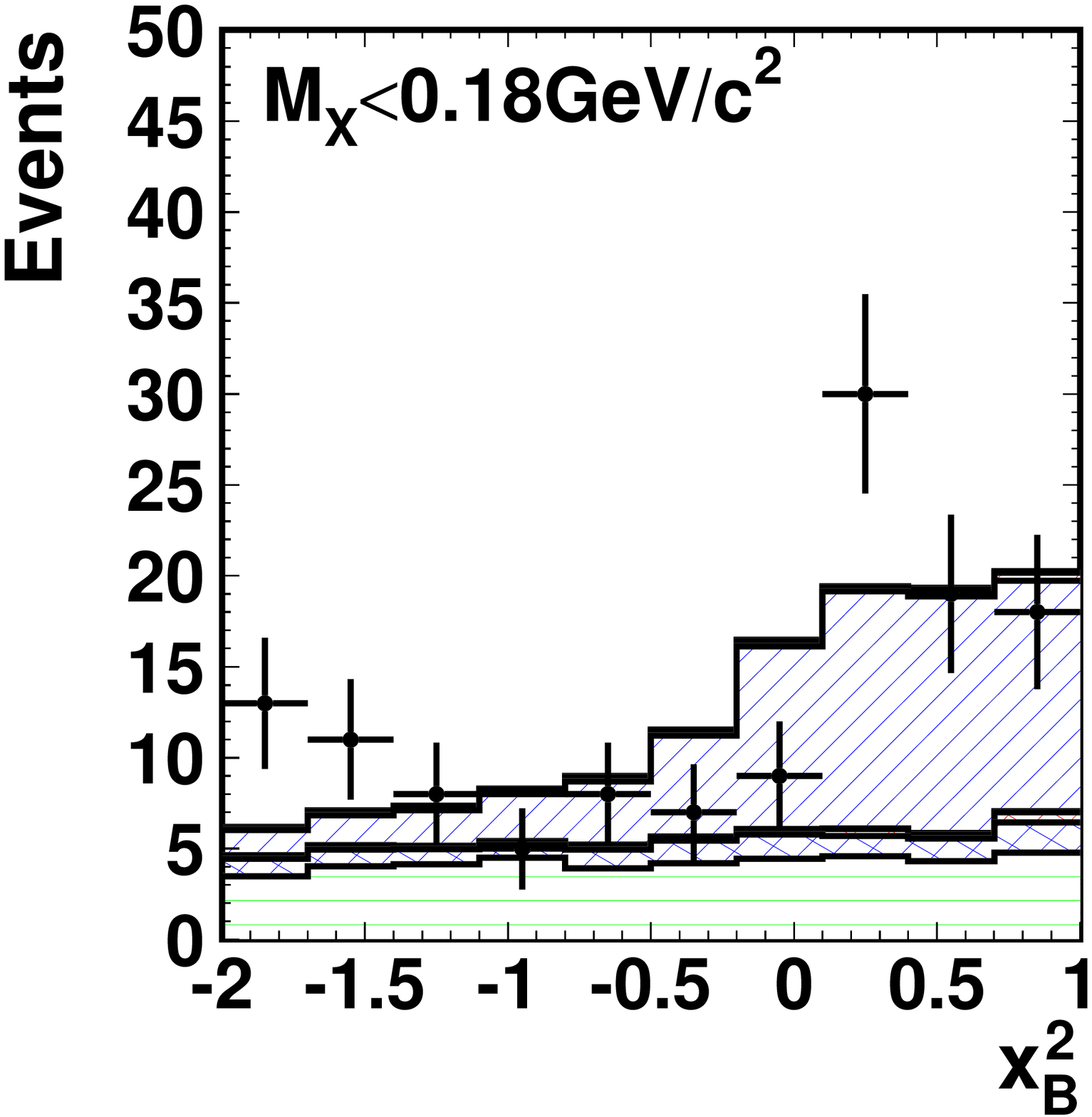,width=1.9in, height=2.0in, angle=0, scale=1.0 }} } &
   \hspace{-0.7cm}{\mbox{\psfig{figure=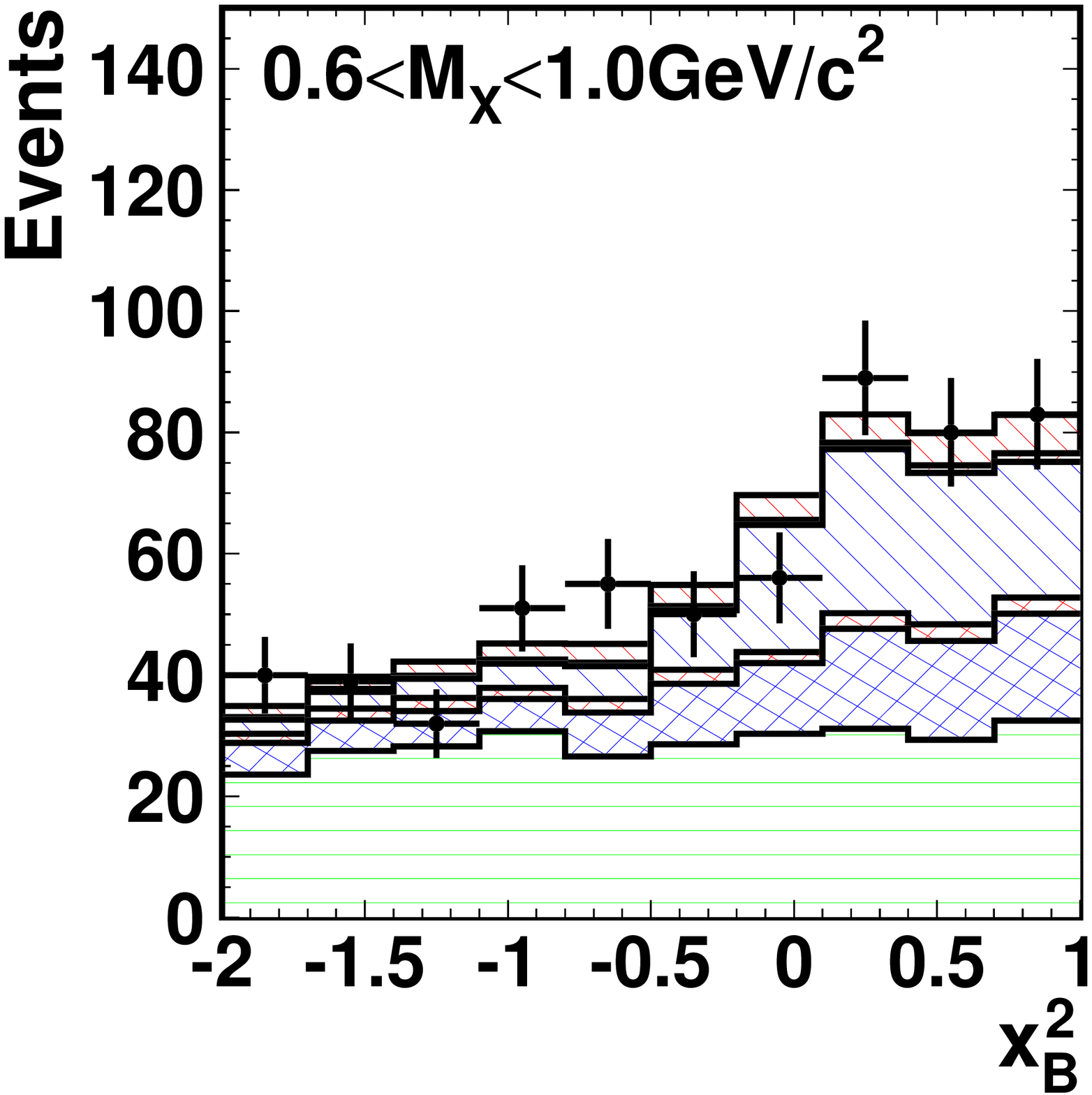,width=1.9in, height=2.0in, angle=0, scale=1.0 } }} \\
  \end{tabular}
  \caption{Projected $M_X$ distribution for $x_B^2>-2$ (left) and $x_B^2$ distributions 
  for the mass region of $\pi$ ($M_X< 0.18 $GeV/$c^2$, middle) 
  and $\rho$ ($ 0.6 < M_X < 1.0 $GeV/$c^2$, right) in all $q^2$ region; points are data.
  Histogram components are $B^0 \to \pi^- \ell^+ \nu$ (red narrow $135^{\circ}$ hatching),
  $B^0 \to \rho^- \ell^+ \nu$ (red wide $45^{\circ}$ hatching), other $X_u \ell^+ \nu$ from $B^0$
  (red cross-hatching) and $B^+ \to \pi^0 \ell^+ \nu$ (blue narrow $45^{\circ}$ hatching),
  $B^+ \to \rho^0 \ell^+ \nu$ (blue wide $135^{\circ}$ hatching), other $X_u \ell^+ \nu$ from $B^+$
  (blue cross-hatching) and $B \bar{B}$ background (green horizontal hatching).}
  \label{fig:fit_allq2}
 \end{center}
\end{figure}

The $B^0 \to \pi^- / \rho^- \ell^+ \nu$ and $B^+ \to \pi^0 / \rho^0 \ell^+ \nu$ signals 
are extracted using binned maximum likelihood fits to the two-dimensional $(x_B^2, M_X)$ distribution,
 where $M_X$ is the nominal pion mass for $B \to \pi \ell^+ \nu$ candidates and 
the invariant mass of two pions for $B \to \rho \ell^+ \nu$ candidates.
The fit includes seven components: the four signal modes 
and the other $B^0 \to X_u^- \ell^+ \nu$ and $B^+ \to X_u^0 \ell^+ \nu$
backgrounds, the background from $B \bar{B}$ events containing no $B \to X_u\ell\nu$.
The PDF (probability density function) for each fit component 
is determined from MC simulation. 
The $\pi / \rho$ signal events exhibit characteristic behavior in 
both the $x_B^2$ and $M_X$ distributions; other $B \to X_u \ell^+ \nu$ 
events exhibit a weak peaking structure in $x_B^2$ but a broad distribution in 
$M_X$; the $B \bar{B}$ background has a relatively flat distribution 
in $x_B^2$ and a broad structure in $M_X$. 
The PDFs in $(x_B^2, M_X)$ for each of the seven fit components are obtained from MC for both $\bar{B}^0$ and $B^-$ tag candidates.
We then fit the two $(x_B^2, M_X)$ distributions for both $\bar{B}^0$ and $B^-$ tags simultaneously;
The fitting is constrained so that the sum of the deduced branching fractions 
for $B \to \pi \ell^+ \nu$, $B \to \rho \ell^+ \nu$ and $B \to$ other 
$X_u \ell^+ \nu$ is equal to the total inclusive branching fraction 
${\mathcal B}(B \to X_u \ell \nu) = (0.25 \pm 0.06)$\%~\cite{BELLE_Kakuno}.
Figure~\ref{fig:fit_allq2} shows the projections on $M_X$ and $x_B^2$ of the 
fitting result for data in the entire $q^2$ region.
The extracted yields for the signal components are 
$N(B^0 \to \pi^- \ell^+ \nu) = 155.8 \pm 20.0$, 
$N(B^0 \to \rho^- \ell^+ \nu) = 92.9 \pm 19.4$,
$N(B^+ \to \pi^0 \ell^+ \nu) = 69.0 \pm 11.4$ and 
$N(B^+ \to \rho^0 \ell^+ \nu) = 135.4 \pm 24.8$,
with the LCSR model used for the four signal PDFs.

%

\begin{figure}[htbp]
 \begin{center}
  \begin{tabular}{ccc}
   \hspace{-0.5cm}{\mbox{\psfig{figure=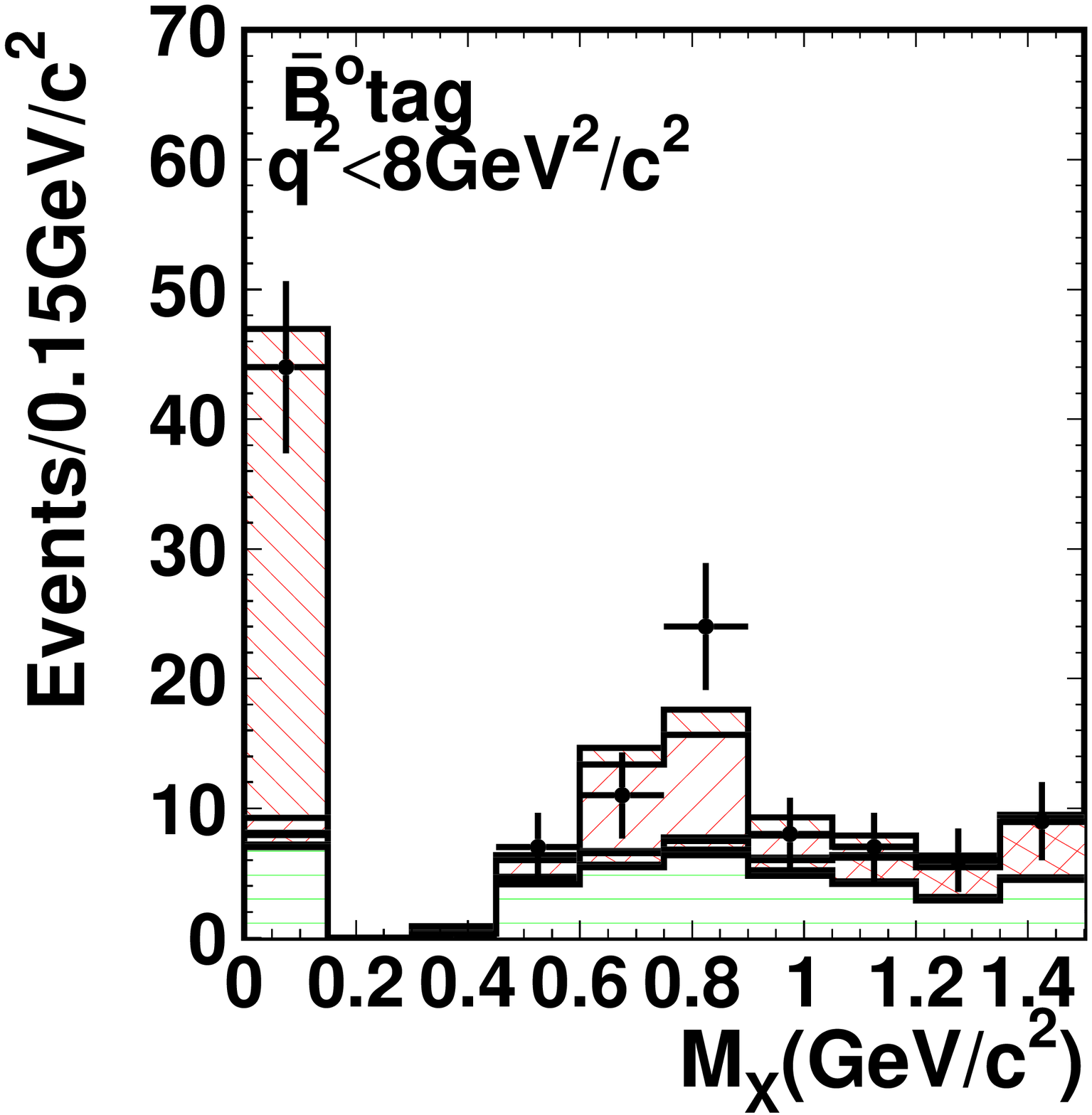,width=1.9in, height=2.0in, angle=0, scale=1.0 }} } &
   \hspace{-0.7cm}{\mbox{\psfig{figure=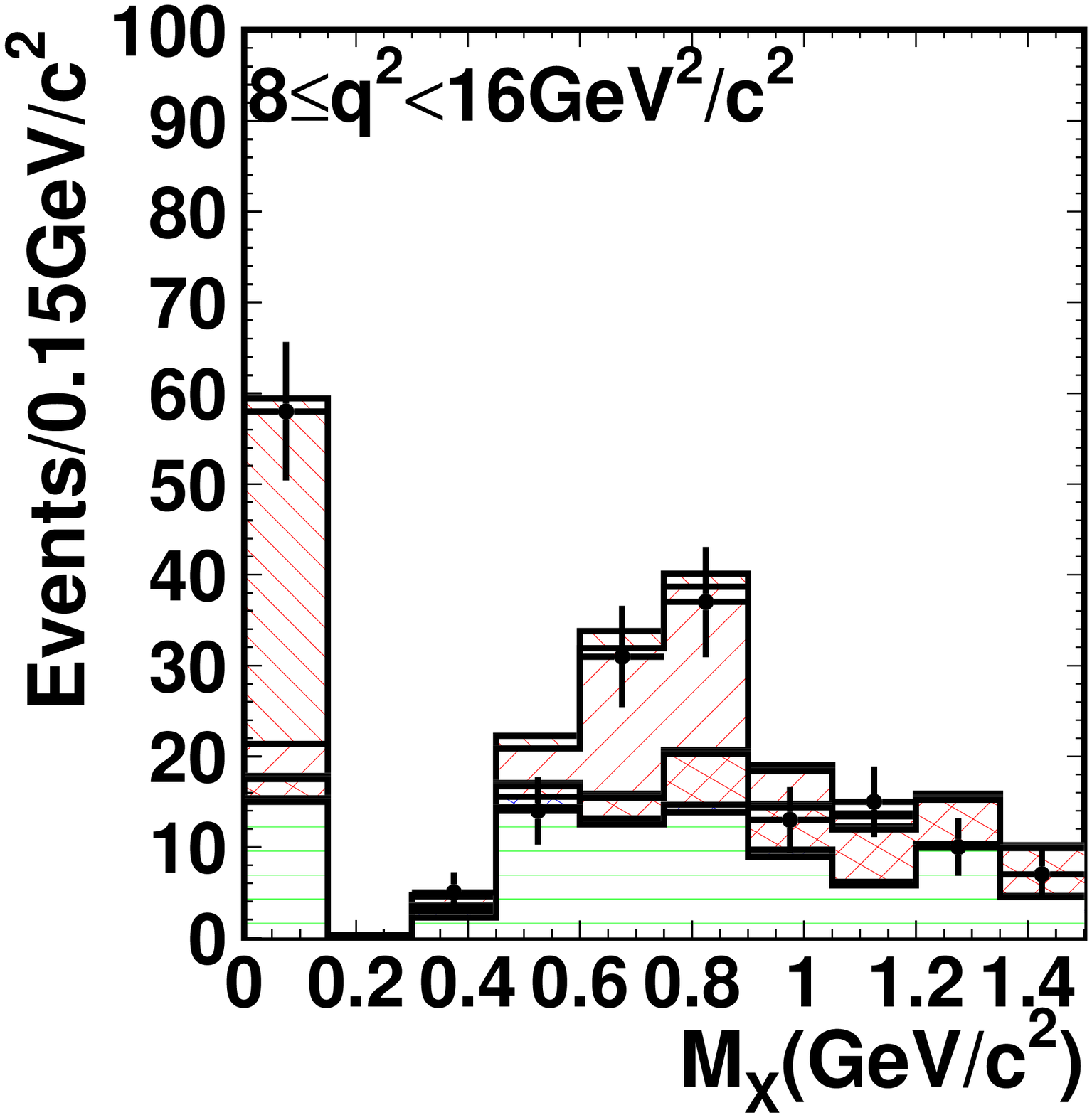,width=1.9in, height=2.0in, angle=0, scale=1.0 }} } &
   \hspace{-0.7cm}{\mbox{\psfig{figure=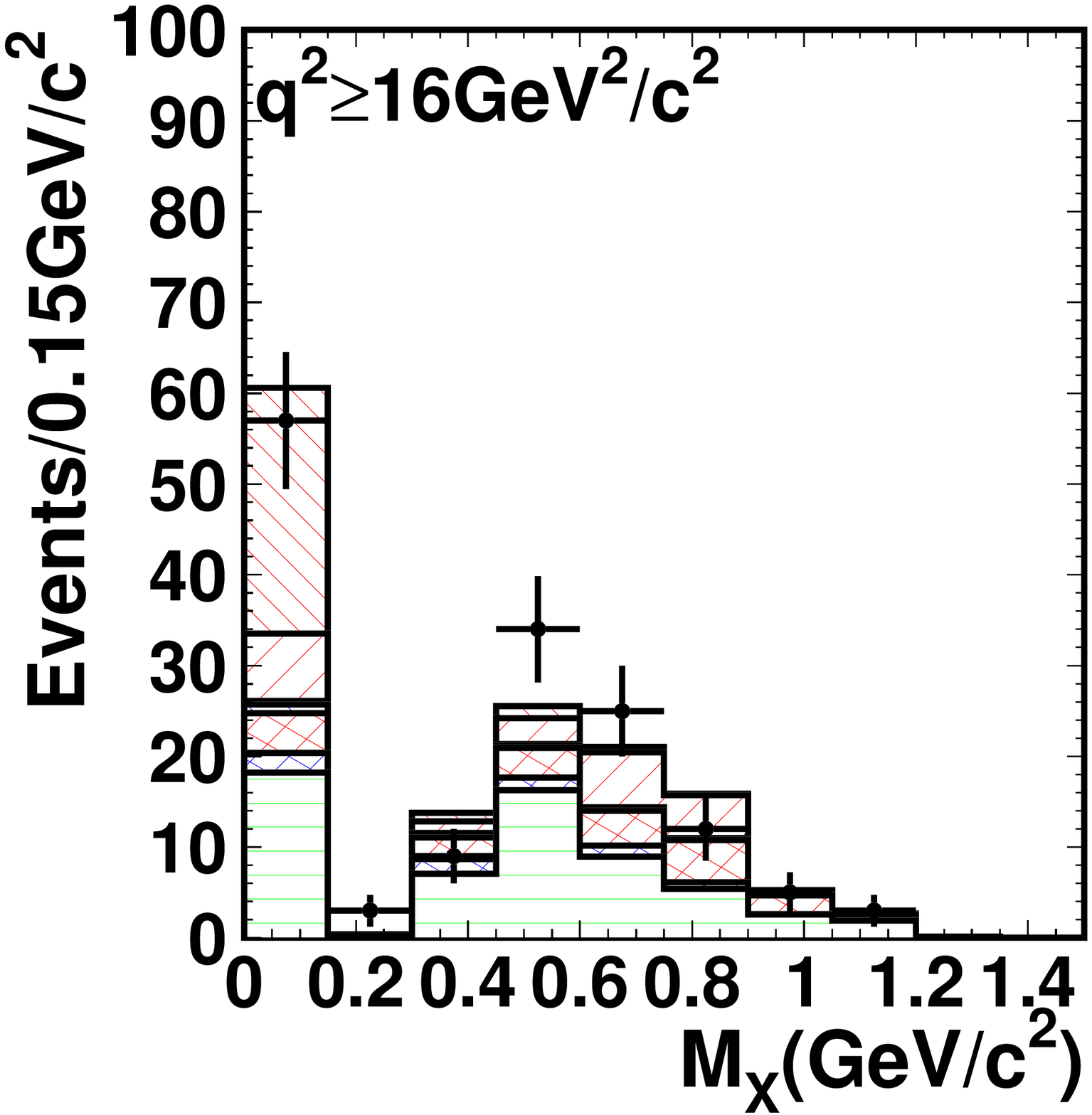,width=1.9in, height=2.0in, angle=0, scale=1.0}} } \\
   \hspace{-0.5cm}{\mbox{\psfig{figure=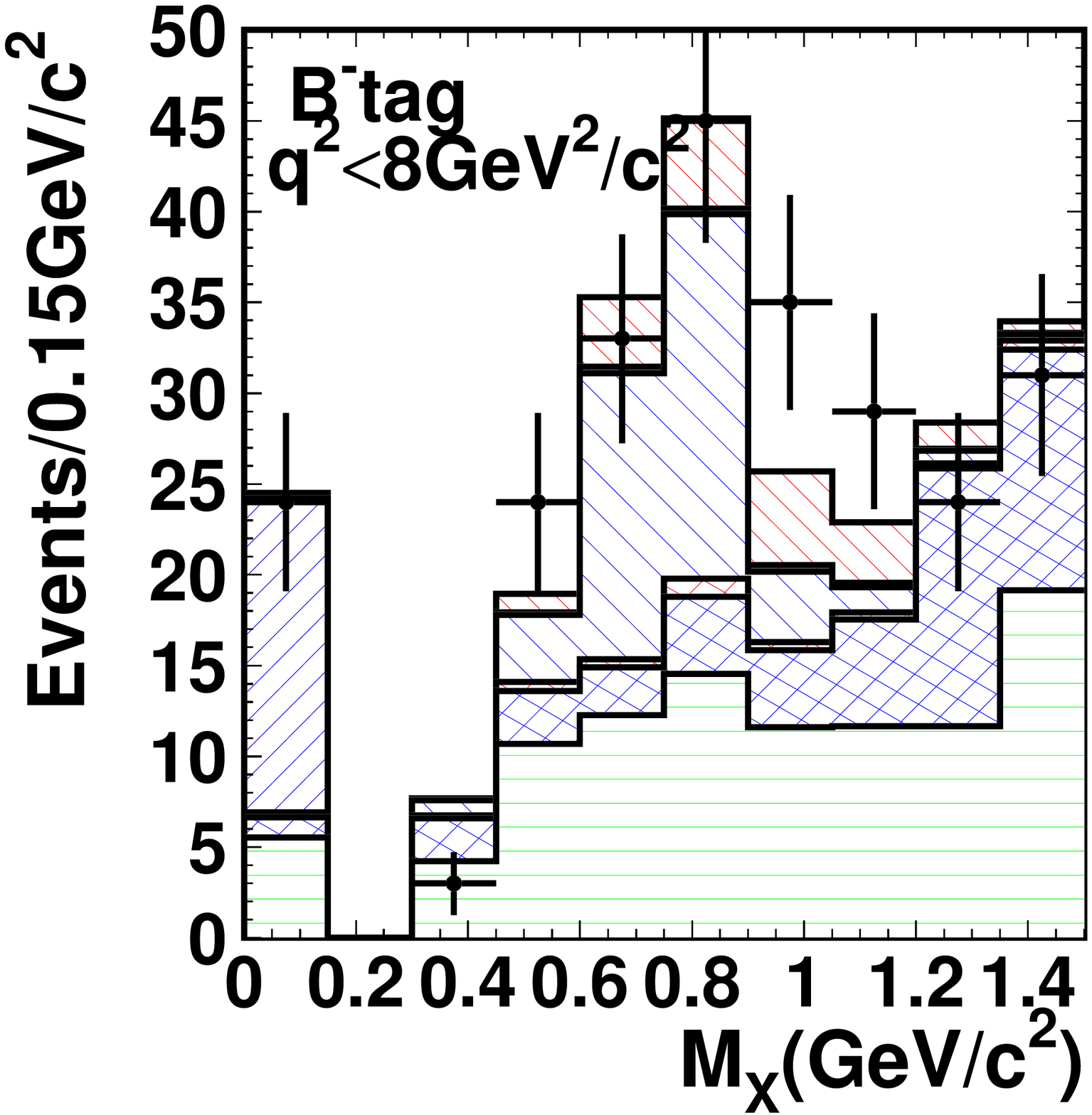,width=1.9in, height=2.0in, angle=0, scale=1.0 }} } &
   \hspace{-0.7cm}{\mbox{\psfig{figure=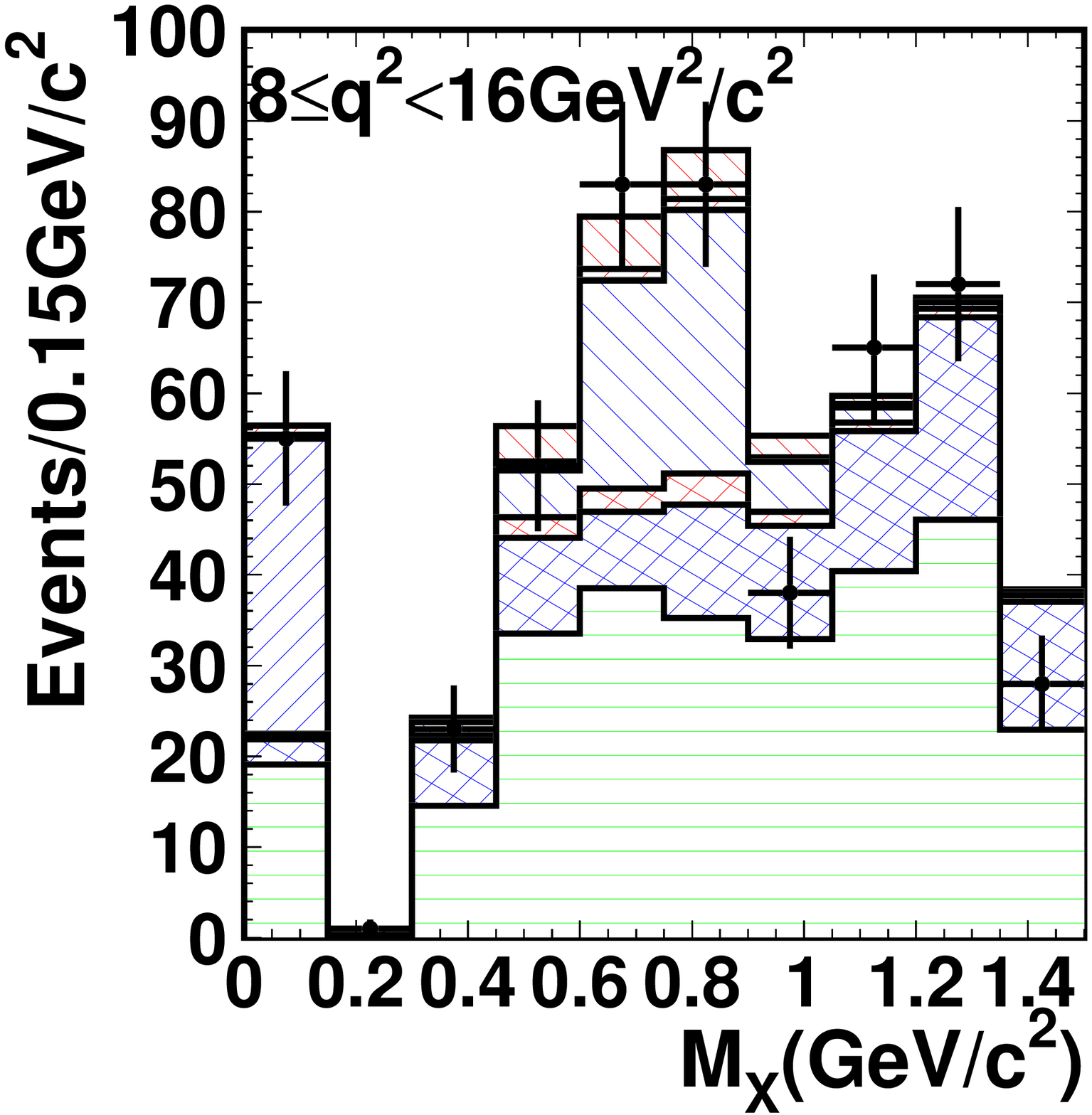,width=1.9in, height=2.0in, angle=0, scale=1.0 }} } &
   \hspace{-0.7cm}{\mbox{\psfig{figure=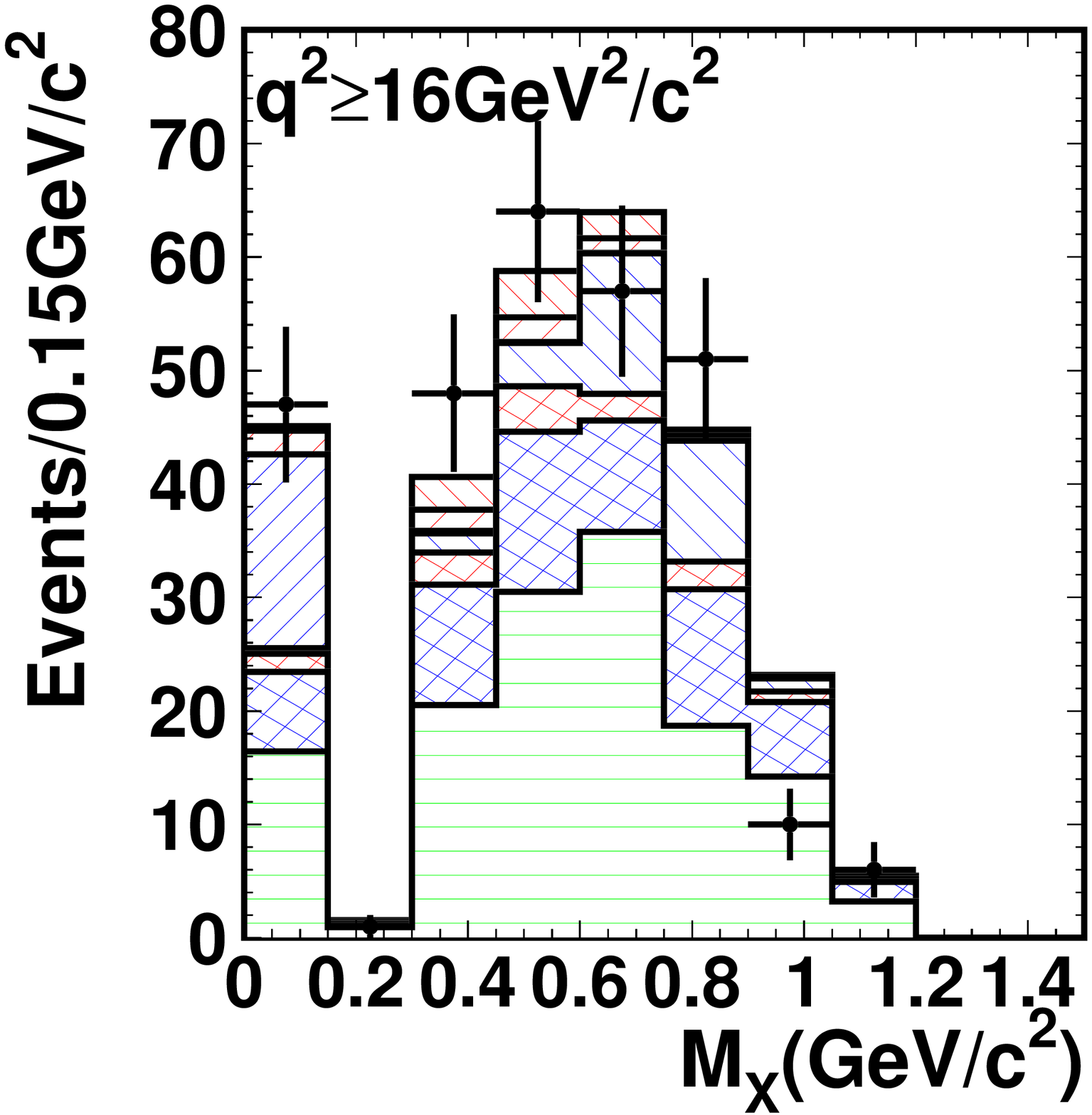,width=1.9in, height=2.0in, angle=0, scale=1.0 } }} \\
  \end{tabular}
  \caption{Projected $M_X$ distribution for $x_B^2>-2$ in each $q^2$ region;
  points are data. Histogram components are $B^0 \to \pi^- \ell^+ \nu$ (red narrow $135^{\circ}$ hatching),
  $B^0 \to \rho^- \ell^+ \nu$ (red wide $45^{\circ}$ hatching), other $X_u \ell^+ \nu$ from $B^0$
  (red cross-hatching) and $B^+ \to \pi^0 \ell^+ \nu$ (blue narrow $45^{\circ}$ hatching),
  $B^+ \to \rho^0 \ell^+ \nu$ (blue wide $135^{\circ}$ hatching), other $X_u \ell^+ \nu$ from $B^+$
  (blue cross-hatching) and $B \bar{B}$ background (green horizontal hatching).}
  \label{fig:fit_subq2}
 \end{center}
\end{figure}

Figure~\ref{fig:fit_subq2} shows projections of the data, separated into three 
$q^2$ bins, $q^2 < 8$ GeV$^2/c^2$, $8 \leq q^2 < 16$ GeV$^2/c^2$ and $q^2 \geq 16$ GeV$^2/c^2$.
Here the normalizations of the other $B \to X_u \ell \nu$ 
and the $B \bar{B}$ background components are fixed to those obtained in the above fitting for the entire $q^2$ region. 
The extracted numbers of events for low/ medium/ high $q^2$ bins are, 
$N(B^0 \to \pi^- \ell^+ \nu) = 64.8 \pm 11.9 ~/~ 63.2 \pm 12.4 ~/~ 40.6 \pm 11.3$,  
$N(B^0 \to \rho^- \ell^+ \nu) = 22.1 \pm 8.0 ~/~ 53.2 \pm 13.5 ~/~ 30.9 \pm 16.0 $, 
$N(B^+ \to \pi^0 \ell^+ \nu) = 18.1 \pm 5.1 ~/~ 34.5 \pm 8.3 ~/~ 18.6 \pm 6.5 $ and 
$N(B^+ \to \rho^0 \ell^+ \nu) = 47.2 \pm 11.2 ~/~ 68.3 \pm 16.5 ~/~ 32.5 \pm 12.3 $.
Table~\ref{tbl:FFdep_q2_Xulnu} summarizes the extracted branching fractions. 
The branching fractions are calculated for each signal FF-model, where we take the average for cross-feed FF-models.
The results are unfolded using the efficiency matrix $\epsilon(q^2_{\rm rec.}, q^2_{\rm true})$ 
for the three $q^2$ intervals prepared for each signal FF-model.
We calculate the total branching fraction by taking the sum of the partial 
branching fractions in the three $q^2$ intervals.

\begin{table}[htbp]
 \begin{center}
  \caption{Extracted branching fractions for each signal mode
    with different FF models in units of 10$^{-4}$: the total branching fraction and the 
    partial branching fractions in three $q^2$ intervals. 
    $\chi^2/n$ and the associated probability for this $\chi^2$ indicate the quality of the fit 
    for the FF shape to the observed $q^2$ distribution.}
  \label{tbl:FFdep_q2_Xulnu}
  {\tabcolsep = 3pt
 \begin{tabular}{l|l|cccccc}
    \hline\hline
Mode & Model    & ${\mathcal B}_{\rm total}$ & ${\mathcal B}_{<8}$ & ${\mathcal B}_{8-16}$ & ${\mathcal B}_{\geq 16}$ & $\chi^2/n$ & $Prob.$\\
\hline
$\pi^- \ell^+ \nu$  & LCSR     & $1.40  \pm 0.19 $ & $0.52  \pm 0.11 $ & $0.51  \pm 0.11 $ & $0.36  \pm 0.10 $ & 0.4/2 & 0.81 \\
                    & ISGW~II  & $1.36  \pm 0.19 $ & $0.52  \pm 0.11 $ & $0.48  \pm 0.11 $ & $0.36  \pm 0.10 $ & 3.6/2 & 0.17 \\
                    & UKQCD    & $1.39  \pm 0.19 $ & $0.53  \pm 0.11 $ & $0.50  \pm 0.11 $ & $0.36  \pm 0.10 $ & 0.2/2 & 0.89 \\
                    & FNAL     & $1.39  \pm 0.19 $ & $0.52  \pm 0.11 $ & $0.51  \pm 0.11 $ & $0.36  \pm 0.10 $ & 0.3/2 & 0.86 \\
                    & HPQCD    & $1.39  \pm 0.19 $ & $0.52  \pm 0.11 $ & $0.50  \pm 0.11 $ & $0.36  \pm 0.10 $ & 0.5/2 & 0.79 \\
                    & Average  & $1.38  \pm 0.19 $ & $0.52  \pm 0.11 $ & $0.50  \pm 0.11 $ & $0.36  \pm 0.10 $ &   --  &  --  \\
\hline
$\rho^- \ell^+ \nu$ & LCSR     & $2.21  \pm 0.54 $ & $0.48  \pm 0.19 $ & $1.18  \pm 0.33 $ & $0.55  \pm 0.39 $ & 2.4/2 & 0.43 \\
                    & ISGW~II  & $2.09  \pm 0.54 $ & $0.44  \pm 0.17 $ & $1.14  \pm 0.32 $ & $0.51  \pm 0.39 $ & 0.5/2 & 0.78 \\
                    & UKQCD    & $2.19  \pm 0.54 $ & $0.48  \pm 0.19 $ & $1.20  \pm 0.33 $ & $0.51  \pm 0.38 $ & 1.7/2 & 0.43 \\
                    & Melikhov & $2.20  \pm 0.54 $ & $0.48  \pm 0.19 $ & $1.22  \pm 0.34 $ & $0.50  \pm 0.37 $ & 1.8/2 & 0.41 \\
                    & Average  & $2.17  \pm 0.54 $ & $0.47  \pm 0.19 $ & $1.19  \pm 0.33 $ & $0.52  \pm 0.38 $ &   --  &  --  \\
\hline
$\pi^0 \ell^+ \nu$  & LCSR     & $0.77  \pm 0.14 $ & $0.19  \pm 0.06 $ & $0.39  \pm 0.10 $ & $0.20  \pm 0.08 $ & 2.9/2 & 0.24 \\
                    & ISGW~II  & $0.77  \pm 0.14 $ & $0.19  \pm 0.06 $ & $0.39  \pm 0.10 $ & $0.20  \pm 0.08 $ & 7.8/2 & 0.02 \\
                    & UKQCD    & $0.77  \pm 0.14 $ & $0.19  \pm 0.06 $ & $0.38  \pm 0.10 $ & $0.20  \pm 0.08 $ & 2.5/2 & 0.28 \\
                    & FNAL     & $0.77  \pm 0.14 $ & $0.19  \pm 0.06 $ & $0.39  \pm 0.10 $ & $0.20  \pm 0.08 $ & 2.8/2 & 0.25 \\
                    & HPQCD    & $0.77  \pm 0.14 $ & $0.19  \pm 0.06 $ & $0.39  \pm 0.10 $ & $0.20  \pm 0.08 $ & 4.8/2 & 0.09 \\
                    & Average  & $0.77  \pm 0.14 $ & $0.19  \pm 0.06 $ & $0.39  \pm 0.10 $ & $0.20  \pm 0.08 $ &   --  &  --  \\
\hline
$\rho^0 \ell^+ \nu$ & LCSR     & $1.36  \pm 0.23 $ & $0.43  \pm 0.11 $ & $0.63  \pm 0.16 $ & $0.29  \pm 0.12 $ & 0.9/2 & 0.64 \\
                    & ISGW~II  & $1.29  \pm 0.22 $ & $0.45  \pm 0.11 $ & $0.57  \pm 0.15 $ & $0.27  \pm 0.12 $ & 1.4/2 & 0.50 \\
                    & UKQCD    & $1.34  \pm 0.23 $ & $0.45  \pm 0.11 $ & $0.62  \pm 0.16 $ & $0.28  \pm 0.12 $ & 0.2/2 & 0.91 \\
                    & Melikhov & $1.35  \pm 0.23 $ & $0.44  \pm 0.11 $ & $0.63  \pm 0.17 $ & $0.28  \pm 0.12 $ & 0.2/2 & 0.92 \\
                    & Average  & $1.33  \pm 0.23 $ & $0.44  \pm 0.11 $ & $0.61  \pm 0.16 $ & $0.28  \pm 0.12 $ &   --  &  --  \\
\hline\hline
  \end{tabular}}
 \end{center}
\end{table}

\section{Systematic Errors}
\label{sec:SystematicErrors}
Tables~\ref{tbl:systematic_pi} and {\ref{tbl:systematic_pi0}} summarize the 
experimental systematic errors on the branching fractions.
The experimental systematic errors can be categorized as originating from uncertainties 
in the signal reconstruction efficiency, the background estimation, and
the normalization.
The total experimental systematic error is the quadratic sum of all 
individual ones.
We also consider the systematic error due to the dependence on the FF model.

The effect from the uncertainty on the signal reconstruction efficiency is evaluated based on the efficiency calibration with the 
$B_{\rm sig} \to D^{*} \ell^+ \nu$ sample, discussed above.
The error is taken to be that on the ratio of observed to expected number of the 
calibration signals (9.3\% for $B^0 \to \pi^-/\rho^- \ell^+ \nu$, 9.2\% for $B \to \pi^0/\rho^0 \ell^+ \nu$).
This gives the largest contribution to the systematic error.
Note that this error is dominated by the statistics of the calibration 
signals, as explained above.
Therefore, accumulation of additional integrated luminosity in the future will help to reduce 
this uncertainty.
We also include residual errors for the reconstruction of the signal 
side: 1\% and 2\% for the detection of each charged and neutral pion, 
respectively, and 2\% for the charged pion selection and 2.1\% for the lepton selection.

The systematic error due to the uncertainty on the inclusive branching  
fraction ${\mathcal B}(B \to X_u \ell \nu)$, which is used to constrain 
$B \to X_u \ell^+ \nu$ background, is estimated by varying this parameter
by its $\pm 1 \sigma$ error.
The uncertainty in the $B \bar{B}$ background shape after our pion multiplicity selection requirements 
($N_{\pi^+}=1$ or $N_{\pi^+}=N_{\pi^0}=1$ for a $\bar{B^0}$ tag and 
 $N_{\pi^0}=1$ or $N_{\pi^+}=N_{\pi^-}=1$ for a $B^-$ tag) is studied in 
the simulation by randomly removing charged tracks and $\pi^0$ according 
to the error in detection efficiency (1\% for a charged track, 2\% for 
$\pi^0$), and also by reassigning identified charged kaons as pions 
according to the uncertainty in the kaon identification efficiency (2\%).
The resultant changes in the extracted branching fractions are assigned
as systematic errors. 
We find a significant uncertainty in the high $q^2$ region ($q^2 > 16$~GeV$^2/c^2$) 
for $B \to \rho \ell^+ \nu$ due to the poor signal-to-noise ratio.
We also vary the fraction of $B \to D^{**} \ell \nu$ decays in the
$B\bar{B}$ background MC by the error quoted in~\cite{PDG2005} to test 
the $B \to X_c \ell \nu$ model dependence in the $B\bar{B}$ background shape.
To assess the uncertainty due to the production rate of $K_L^0$, we vary
the production rate in the MC simulation by the uncertainty in the inclusive branching fraction for 
$B \to K^0 ~X$ quoted in~\cite{PDG2005}.
  
For the normalization, we consider the uncertainty in the number of 
$B^0 \bar{B^0}$ and $B^+ B^-$ pairs: 
the ratio of $B^+B^-$ to $B^0 \bar{B^0}$ pairs ($f_+/f_0$), $f_+/f_0 = 1.029\pm 0.035$ ~\cite{HFAG05}
, the mixing parameter ($\chi_d$), $\chi_d = 0.186\pm 0.004$ ~\cite{PDG2005}, and the measured
number of $B \bar{B}$ pairs ($N_{B\bar{B}}$, 1.1\%).
\begin{table}[htbp]
 \begin{center}
  \caption{Summary of systematic errors (\%) for ${\mathcal B}(B^0 \to \pi^- / \rho^- \ell^+ \nu)$.}
  \label{tbl:systematic_pi}

  {\tabcolsep = 2pt \footnotesize
  \begin{tabular}{c|ccc|c|c|ccc|c|c}
   \hline\hline
          & \multicolumn{5}{c|}{$B^0 \to \pi^{-} \ell^{+} \nu$} & \multicolumn{5}{c}{$B^0 \to \rho^{-} \ell^{+} \nu$}            \\
          & \multicolumn{5}{c|}{$q^2$ interval (GeV$^2/c^2$)} & \multicolumn{5}{c}{$q^2$ interval (GeV$^2/c^2$)} \\
   Source &  $q^2<8$ &  $ 8-16$ &  $\geq16$ &  $< 16$ &  all &  $q^2<8$ &  $8-16$ &  $\geq16$ &  $< 16$ &  all     \\
   \hline
   Tracking efficiency        &      1  &    1  &    1  &    1  &    1  &     1  &    1  &    1  &    1  &    1  \\ 
   $\pi^{0}$ reconstruction   &     --  &   --  &   --  &   --  &   --  &     2  &    2  &    2  &    2  &    2  \\ 
   Lepton identification      &    2.1  &  2.1  &  2.1  &  2.1  &  2.1  &   2.1  &  2.1  &  2.1  &  2.1  &  2.1  \\ 
   Kaon identification        &      2  &    2  &    2  &    2  &    2  &     2  &    2  &    2  &    2  &    2  \\ 
   $D^* \ell \nu$ calibration &    9.3  &  9.3  &  9.3  &  9.3  &  9.3  &   9.3  &  9.3  &  9.3  &  9.3  &  9.3  \\ 
   $Br(X_u \ell \nu)$ in the fitting        
                              &    0.8  &  2.4  &  1.8  &  1.6  &  1.4  &   4.8  &  3.9  & 23.8  &  1.9  &  7.1  \\ 
   $B\bar{B}$ background shape
                              &    1.1  &  2.2  &  2.8  &  1.2  &  1.3  &   3.8  &  2.9  & 17.0  &  3.0  &  6.1  \\
   $Br(D^{**} \ell \nu)$      &    1.0  &  1.5  &  0.2  &  1.2  &  0.9  &   0.5  &  0.3  &  2.5  &  0.3  &  0.8  \\
   $K_L^0$ production rate    &    0.1  &  0.3  &  0.4  &  0.2  &  0.3  &   1.0  &  0.7  &  2.9  &  0.8  &  1.3  \\
   $N_{B\overline{B}}$        &    1.1  &  1.1  &  1.1  &  1.1  &  1.1  &   1.1  &  1.1  &  1.1  &  1.1  &  1.1  \\ 
   $\it{f}_+ / \it{f}_0$      &    1.7  &  1.7  &  1.7  &  1.7  &  1.7  &   1.7  &  1.7  &  1.7  &  1.7  &  1.7  \\ 
   $\chi_d$                   &    2.2  &  2.2  &  2.2  &  2.2  &  2.2  &   2.2  &  2.2  &  2.2  &  2.2  &  2.2  \\ 
   \hline
   exp. total                 &   10.4  & 10.9 & 10.8  &  10.5 &  10.5  &  12.1  & 11.5  & 31.3  & 11.1  & 14.1  \\ 
   \hline\hline
   FF for signal             &    0.7  &  3.8  &  0.9  &   2.2 &  1.8  &   6.1  &  3.5  &  6.8  &  4.3  &  3.6  \\ 
   FF for cross-feed         &    1.8  &  2.1  &  1.5  &   1.9 &  1.4  &   0.5  &  0.7  &  2.4  &  0.6  &  1.0  \\ 
   \hline						          	  			      	   	     
   FF total                  &    1.9  &  4.3  &  1.7  &   2.9 &  2.3  &   6.1  &  3.6  &  7.2  &  4.3  &  3.7  \\
   \hline\hline
  \end{tabular}}
 \end{center}
\end{table}

\begin{table}[htbp]
 \begin{center}
  \caption{Summary of systematic errors (\%) for ${\mathcal B}(B^+ \to \pi^0 / \rho^0 \ell^+ \nu)$.}
  \label{tbl:systematic_pi0}

  {\tabcolsep = 2pt \footnotesize
  \begin{tabular}{c|ccc|c|c|ccc|c|c}
   \hline\hline
          & \multicolumn{5}{c|}{$B^+ \to \pi^0 \ell^{+} \nu$} & \multicolumn{5}{c}{$B^+ \to \rho^0 \ell^{+} \nu$}            \\
          & \multicolumn{5}{c|}{$q^2$ interval (GeV$^2/c^2$)} & \multicolumn{5}{c}{$q^2$ interval (~GeV$^2/c^2$)} \\
   Source &  $q^2<8$ &  $ 8-16$ &  $\geq16$ &  $< 16$ &  all &  $q^2<8$ &  $8-16$ &  $\geq16$ &  $< 16$ &  all     \\
   \hline
   Tracking efficiency        &     --  &   --  &   --  &   --  &   --  &     2  &    2  &    2  &    2  &    2  \\ 
   $\pi^{0}$ reconstruction   &      2  &    2  &    2  &    2  &    2  &    --  &   --  &   --  &   --  &   --  \\ 
   Lepton identification      &    2.1  &  2.1  &  2.1  &  2.1  &  2.1  &   2.1  &  2.1  &  2.1  &  2.1  &  2.1  \\ 
   Kaon identification        &     --  &   --  &   --  &   --  &   --  &     4  &    4  &    4  &    4  &    4  \\ 
   $D^* \ell \nu$ calibration &    9.2  &  9.2  &  9.2  &  9.2  &  9.2  &   9.2  &  9.2  &  9.2  &  9.2  &  9.2  \\ 
   $Br(X_u \ell \nu)$ in the fitting        
                              &    0.2  &  3.1  &  3.0  &  2.1  &  1.2  &   2.0  &  3.7  & 20.0  &  3.0  &  6.6  \\ 
   $B\bar{B}$ background shape
                              &    1.9  &  5.5  &  2.7  &  4.3  &  3.7  &   5.3  &  4.3  & 16.3  &  1.5  &  2.8  \\
   $Br(D^{**} \ell \nu)$      &    1.3  &  0.8  &  0.8  &  0.9  &  0.9  &   0.2  &  1.6  &  3.0  &  0.9  &  1.4  \\
   $K_L^0$ production rate    &    0.3  &  1.1  &  0.6  &  0.8  &  0.8  &   0.3  &  0.2  &  1.9  &  0.1  &  0.4  \\
   $N_{B\overline{B}}$        &    1.1  &  1.1  &  1.1  &  1.1  &  1.1  &   1.1  &  1.1  &  1.1  &  1.1  &  1.1  \\ 
   $\it{f}_+ / \it{f}_0$      &    1.7  &  1.7  &  1.7  &  1.7  &  1.7  &   1.7  &  1.7  &  1.7  &  1.7  &  1.7  \\ 
   \hline
   exp. total                 &   10.1  & 11.8  & 10.7  & 11.0  & 10.7  &  12.1  & 12.2  & 28.1  & 11.2  & 12.9  \\ 
   \hline\hline
   FF for signal             &    1.2  &  0.5  &  1.3  &  0.3  &  0.2  &   2.1  &  7.1  &  3.9  &  3.7  &  3.5  \\ 
   FF for cross-feed         &    0.7  &  0.8  &  0.6  &  0.8  &  0.6  &   3.3  &  1.1  &  1.0  &  1.5  &  1.2  \\ 
   \hline						          	  			      	   	     
   FF total                  &    1.4  &  0.9  &  1.4  &  0.8  &  0.6  &   3.9  &  7.2  &  4.0  &  4.0  &  3.7  \\
   \hline\hline
  \end{tabular}}
 \end{center}
\end{table}

The dependence of the extracted branching fractions on the FF model has been 
studied by repeating the above fitting procedure with various FF 
models for the signal mode and also for the cross-feed mode 
($B \to \pi \ell \nu \leftrightarrow B \to \rho \ell \nu$).
We consider the models listed in Table~\ref{tbl:FFdep_q2_Xulnu}.
For the extracted ${\mathcal B}(B \to \pi^- \ell^+ \nu ~(\pi^0 \ell^+ \nu))$,
the standard deviation among the models is $<1.7 ~(0.9)$\% for 
$B \to \pi \ell^+ \nu$  and $<1.9 ~(0.5)$\% for  
$B \to \rho \ell^+ \nu$.
For ${\mathcal B}(B \to \rho^- \ell^+ \nu ~(\rho^0 \ell^+ \nu))$, the standard deviation is
$<2.9 ~(3.6)$\% for 
$B \to \rho \ell^+ \nu$ and $<1.0 ~(1.3)$\% for  
$B \to \pi \ell^+ \nu$.
The total error due to FF model dependence is the
quadratic sum of the maximum variations with the signal and cross-feed 
FF models.

\section{Results}
Table~\ref{tbl:summary_br} summarizes our measurements of the total and partial 
branching fractions for the four signal modes.
Each branching fraction is obtained by taking the simple average of 
the values obtained from the FF models shown in 
Table~\ref{tbl:FFdep_q2_Xulnu}.
The errors shown in the table are statistical, experimental systematic,
and model dependence due to form-factor uncertainties.
The obtained branching fractions for $B^0 \to \pi^-/\rho^- \ell^+ \nu$
are consistent with the existing measurements by CLEO~\cite{CLEO2003} 
and BaBar~\cite{BABAR2005}.
The overall uncertainty on our result for $B^0 \to \pi^- \ell^+ \nu$ (17\%) is 
comparable to those on CLEO and BaBar results based on $\nu$-reconstruction.
Our results for $B^0 \to \rho^- \ell^+ \nu$ have the smallest uncertainty.

\begin{table}[htbp]
 \begin{center}
  \caption{Summary of the obtained branching fractions.
The errors are statistical, experimental systematic, and systematic due 
to form-factor uncertainties.}
  \label{tbl:summary_br}
  \begin{tabular}{cc|c}
   \hline\hline
   Modes    &   $q^2$ region (GeV$^2/c^2)$  & Branching fraction ($\times 10^{-4}$)\\
   \hline
   $B^0 \to \pi^- \ell^+ \nu$   & ~Total    & $1.38 \pm 0.19 \pm 0.14 \pm 0.03$ \\
                                & ~$ > 16$  & $0.36 \pm 0.10 \pm 0.04 \pm 0.01$ \\
                                & ~$ < 16$  & $1.02 \pm 0.16 \pm 0.11 \pm 0.03$ \\
   \hline
   $B^+ \to \pi^0 \ell^+ \nu$   & ~Total    & $0.77 \pm 0.14 \pm 0.08 \pm 0.00 $ \\
                                & ~$ > 16$  & $0.20 \pm 0.08 \pm 0.02 \pm 0.00 $ \\
                                & ~$ < 16$  & $0.57 \pm 0.12 \pm 0.06 \pm 0.00 $ \\
   \hline
   $B^0 \to \rho^- \ell^+ \nu$  & ~Total    & $2.17 \pm 0.54 \pm 0.31 \pm 0.08$ \\   
   \hline
   $B^+ \to \rho^0 \ell^+ \nu$  & ~Total    & $1.33 \pm 0.23 \pm 0.17 \pm 0.05$ \\   
   \hline\hline
  \end{tabular}
 \end{center}
\end{table}

\begin{figure}[htbp]
 \begin{center}
  \begin{tabular}{cc}
   \hspace{-0.5cm}{\mbox{\psfig{figure=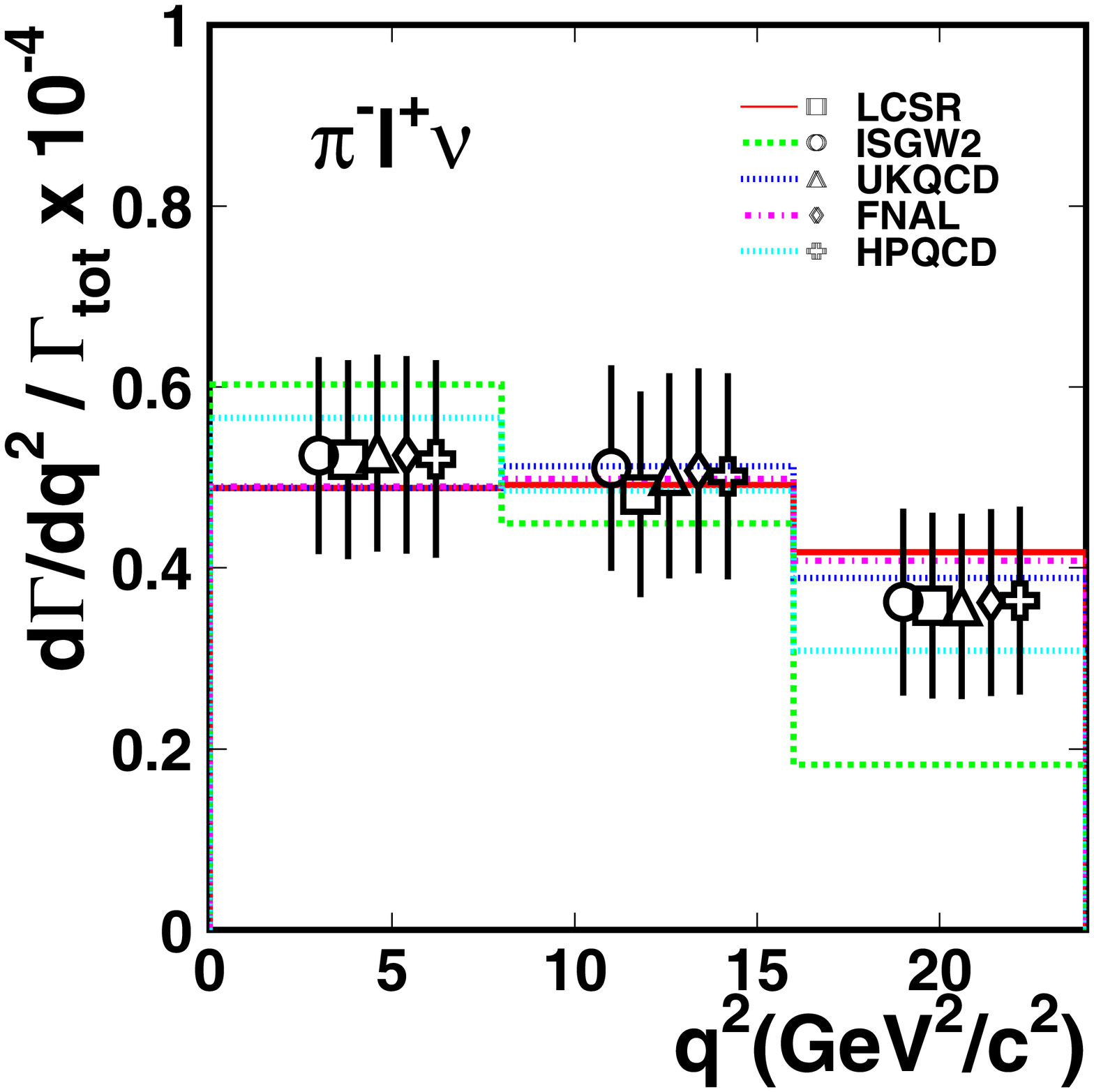,width=2.5in, height=2.5in, angle=0, scale=1.1 } }} &
   \hspace{-0.5cm}{\mbox{\psfig{figure=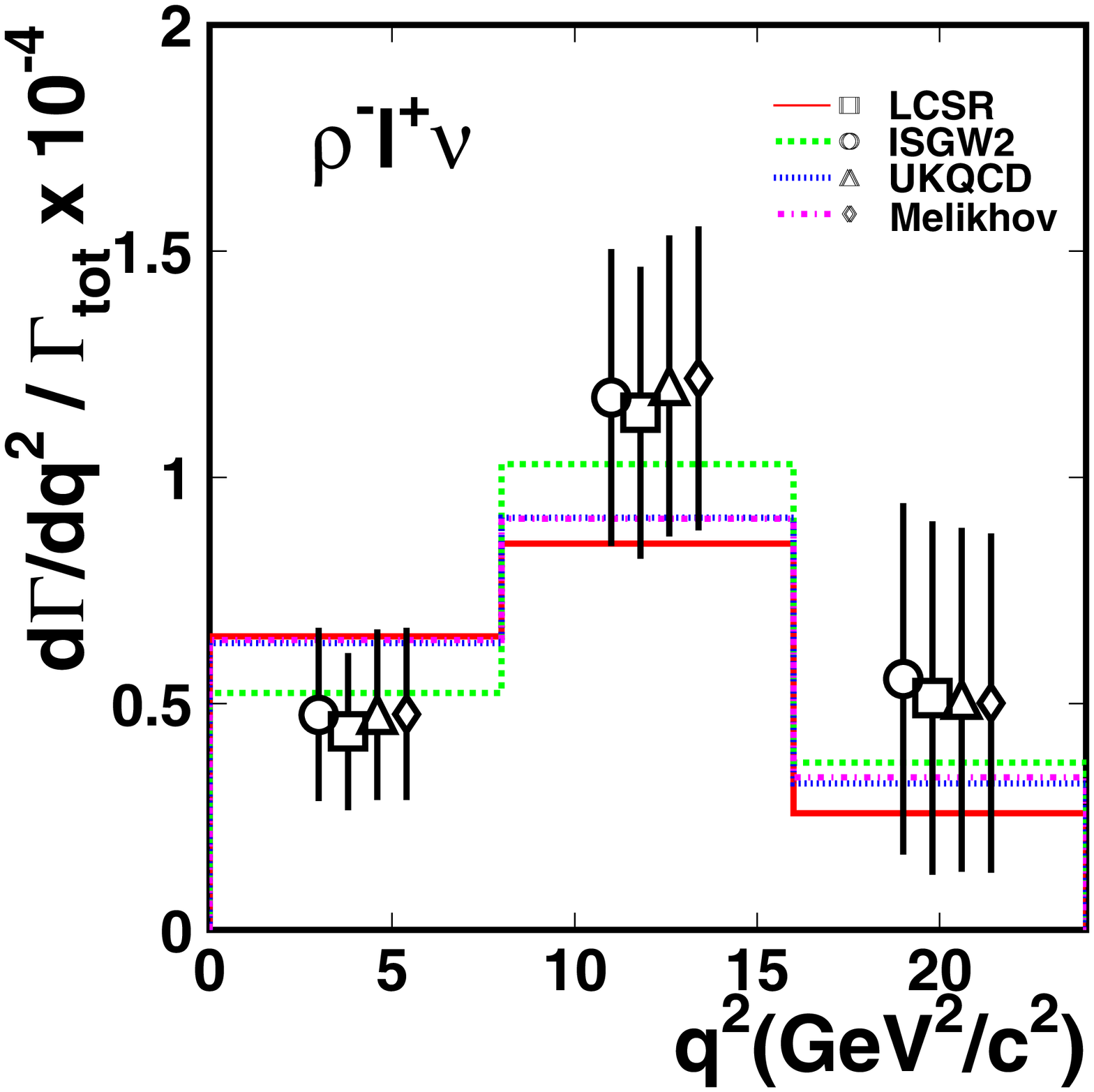,width=2.5in, height=2.5in, angle=0, scale=1.1 } }}
   \vspace{0.0cm} \\
   \hspace{-0.5cm}{\mbox{\psfig{figure=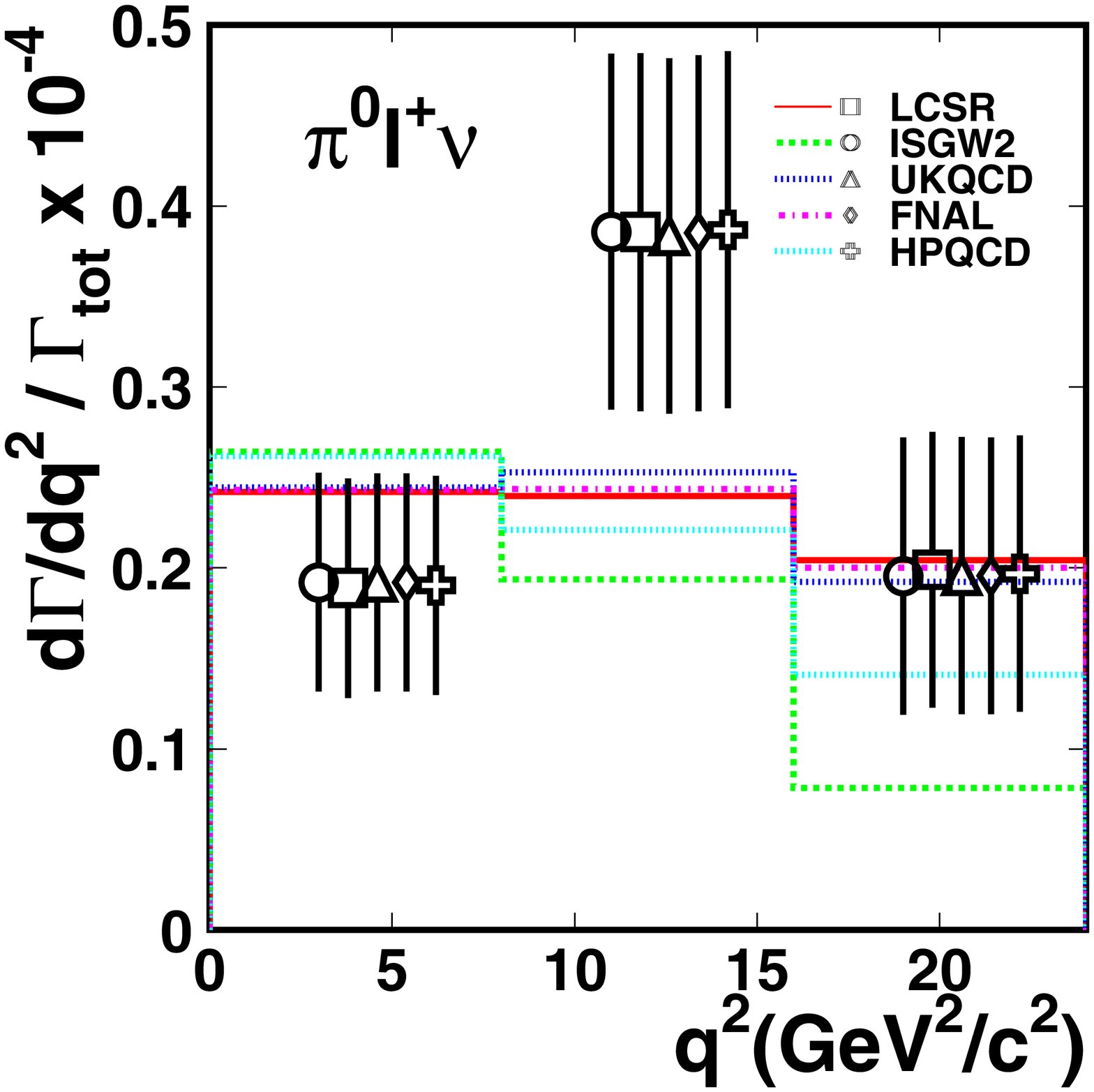,width=2.5in, height=2.5in, angle=0, scale=1.1 } }} &
   \hspace{-0.5cm}{\mbox{\psfig{figure=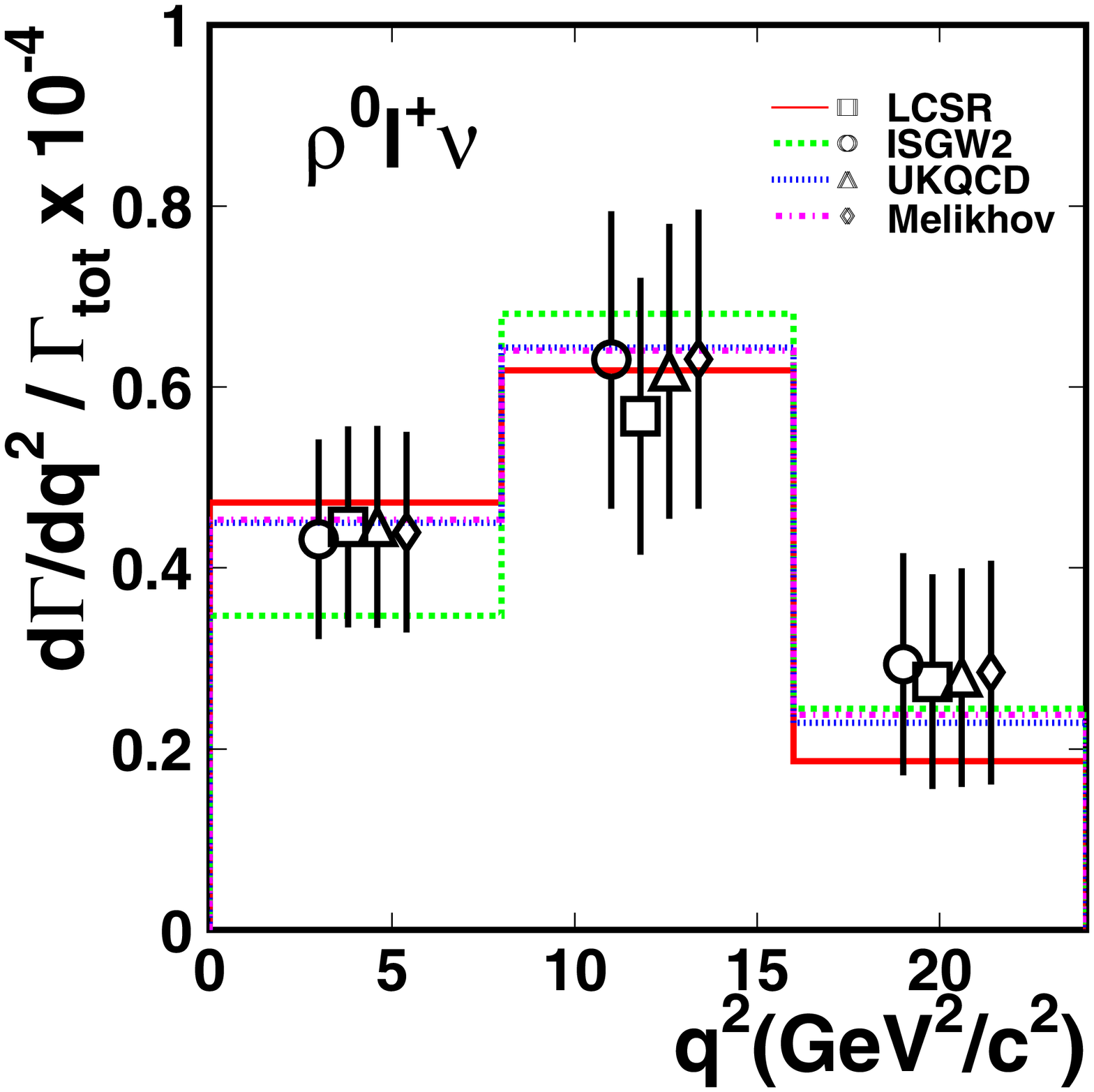,width=2.5in, height=2.5in, angle=0, scale=1.1 } }}\\

  \end{tabular}
  \caption{Extracted $q^2$ distribution. Data points are shown for different FF
    models used to estimate the detection efficiency. 
    Lines are for the best fit of the FF shapes to the obtained $q^2$ 
    distribution.}
  \label{fig:q2dist}
 \end{center}
\end{figure}

Figure~\ref{fig:q2dist} presents the measured $q^2$ distributions for 
each signal mode, overlaid with the best fits of FF shapes to the data.
To be self-consistent, the shape of a particular FF model is fit to the 
$q^2$ distribution extracted with the same FF model.
The quality of the fit in terms of $\chi^2$ and the probability of 
$\chi^2$, shown in Table~\ref{tbl:FFdep_q2_Xulnu}, may provide one way 
to discriminate among the models. 
From our results, the ISGW II model is disfavored for $B \to \pi \ell^+ \nu$.

In this work, the $B^0 \to \pi^- \ell^+ \nu / B^+ \to \pi^0 \ell^+ \nu$
and $B^0 \to \rho^- \ell^+ \nu / B^+ \to \rho^0 \ell^+ \nu$ signals
are extracted separately, which allows us to test the isospin relations.
From the obtained branching fractions and the $B$ meson lifetimes in
~\cite{PDG2005}, the ratios of decay rates are found to be,
\begin{eqnarray}
\frac{\Gamma(B^0 \to \pi^- \ell^+ \nu)}{\Gamma(B^+ \to \pi^0 \ell^+ \nu)}
= (1.92 \pm 0.43 \pm 0.28),\\
\frac{\Gamma(B^0 \to \rho^- \ell^+ \nu)}{\Gamma(B^+ \to \rho^0 \ell^+ \nu)}
= (1.74 \pm 0.53 \pm 0.33),
\end{eqnarray}  
where the first and second errors are statistical and systematic errors,
respectively.
Both ratios are found to be consistent with the isospin relations;
$\Gamma(B^0 \to \pi^-(\rho^-) \ell^+ \nu) = 
2\Gamma(B^+ \to \pi^0(\rho^0) \ell^+ \nu)$. 

%
%
The obtained branching fractions in Table~\ref{tbl:summary_br}
can be used to extract $|V_{ub}|$ using the relation,
\begin{equation}
|V_{ub}| = \sqrt{ \frac{{\mathcal B}(B \to \pi \ell^+ \nu) }
  {{\tilde \Gamma}_{thy}~\tau_{B} } },
\end{equation}
where ${\tilde \Gamma}_{thy}$ is the form-factor normalization, 
predicted from theories.
In Table~\ref{tbl:summary_br}, we list the partial branching fractions for $B \to \pi \ell^+ \nu$
decays in the $q^2$ region above 16 GeV$^2/c^2$, where the LQCD 
calculations are most reliable.
The table provides also the results in the region below 
16 GeV$^2/c^2$, so that one can deduce $|V_{ub}|$ based on other
approaches such as LCSR calculations~\cite{Ball04_pi,Ball04_rho}.

In this paper we calculate $|V_{ub}|$ based on the $B \to \pi \ell^+ \nu$ data 
in the high $q^2$ region and the form factor predicted by recent 
unquenched LQCD calculations. 
Their predictions (${\tilde \Gamma}_{thy}$) for the $q^2 \geq 16$\,GeV$^2/c^2$ 
region are ${\tilde \Gamma}_{thy}(B^0 \to \pi^- \ell^+ \nu) = 1.83 \pm 0.50$ ps$^{-1}$ (FNAL)~\cite{FNAL04} and 
${\tilde \Gamma}_{thy}(B^0 \to \pi^- \ell^+ \nu) = 1.46 \pm 0.35$ ps$^{-1}$ (HPQCD)~\cite{HPQCD04}.
We use $\tau_{B^0} = 1.532\pm 0.009$ ps and $\tau_{B^+} = 1.638\pm 0.011$ ps 
~\cite{PDG2005}, and we use isospin symmetry to relate 
${\tilde \Gamma}_{thy}$ for $B^0 \to \pi^-$ and $B^+ \to \pi^0$ transitions. 
The results for $B^0 \to \pi^- \ell^+ \nu$ and $B^+ \to \pi^0 \ell^+ \nu$ 
are then averaged, weighted by their respective statistical errors.

\begin{table}[htbp]
 \begin{center}
  \caption{Summary of $|V_{ub}|$ obtained from the $B \to \pi \ell^+ \nu$
data in the $q^2 \ge 16$\,GeV$^2/c^2$ region. 
The first and second errors are experimental statistical and systematic errors, 
respectively. The third error stems from the error on
${\tilde \Gamma}_{thy}$ quoted by the LQCD authors.}
  \label{tbl:summary_vub}
  \begin{tabular}{cccc}
   \hline\hline
  Theory  &  ${\tilde \Gamma}_{thy}$(ps$^{-1}$) & Mode & $|V_{ub}| (\times 10^{-3})$ \\
   \hline
  FNAL    &  $1.83 \pm 0.50$ & $\pi^- \ell^+ \nu$                       & $3.59 \pm 0.51 \pm 0.20 ^{+0.62}_{-0.41}$ \\
          &                  & $\pi^0 \ell^+ \nu$                       & $3.63 \pm 0.70 \pm 0.20 ^{+0.63}_{-0.41}$ \\
          &                  & $\pi^- \ell^+ \nu + \pi^0 \ell^+ \nu$    & $3.60 \pm 0.41 \pm 0.20 ^{+0.62}_{-0.41}$ \\
   \hline
  HPQCD   &  $1.46 \pm 0.35$ & $\pi^- \ell^+ \nu$                       & $4.02 \pm 0.57 \pm 0.22 ^{+0.59}_{-0.41}$ \\
          &                  & $\pi^0 \ell^+ \nu$                       & $4.06 \pm 0.78 \pm 0.22 ^{+0.60}_{-0.41}$ \\
          &                  & $\pi^- \ell^+ \nu + \pi^0 \ell^+ \nu$    & $4.03 \pm 0.46 \pm 0.22 ^{+0.59}_{-0.41}$ \\
   \hline\hline
  \end{tabular}
 \end{center}
\end{table}

Table~\ref{tbl:summary_vub} summarizes the results, where the first and
second errors are the experimental statistical and systematic errors,
respectively.
The third error is based on the error on ${\tilde \Gamma}_{thy}$ quoted by the LQCD authors.
These theoretical errors are asymmetric because we assign them by taking the variation 
in $|V_{ub}|$ when ${\tilde \Gamma}_{thy}$ is varied by the quoted errors.  
The values are in agreement with those from inclusive $B\to X_u\ell\nu$ decays~\cite{HFAG_LP05}.

To summarize, we have measured the branching fractions of the decays 
$B\to \pi\ell\nu$ and $B\to\rho\ell\nu$ in $2.75\times 10^8$ $B\bar 
B$ events using a method which tags one $B$ in the mode $B\to 
D^{(*)}\ell\nu$.
Our results are consistent with previous measurements, and their 
precision is comparable to that of results from other experiments.
The ratios of results for neutral and charged $B$ meson modes are found 
to be consistent with isospin.
The partial rates are measured in three bins of $q^2$ and compared 
with  distributions predicted by several theories.
 From the rate in the region $q^2 \ge 16$~GeV$^2/c^2$ and recent 
results from LQCD calculations, we extract $|V_{ub}|$:
\begin{eqnarray}
|V_{ub}|_{(q^2 \geq 16 ~{\rm GeV}^2/c^2)}^{\pi^- \ell^+ \nu + \pi^0 \ell^+ \nu } & = &
(3.60 \pm 0.41 \pm 0.20 ^{+0.62}_{-0.41}) \times 10^{-3} (\mbox{FNAL LQCD}), \\
|V_{ub}|_{(q^2 \geq 16 ~{\rm GeV}^2/c^2)}^{\pi^- \ell^+ \nu + \pi^0 \ell^+ \nu } & = &
(4.03 \pm 0.46 \pm 0.22 ^{+0.59}_{-0.41}) \times 10^{-3} (\mbox{HPQCD LQCD}).
\end{eqnarray}

The experimental precision on these values is 13\%, currently 
dominated by the statistical error of 11\%.  By accumulating more 
integrated luminosity, a measurement with errors below 10\% is 
feasible.
With improvements to unquenched LQCD calculations, the present method 
may provide a precise determination of $|V_{ub}|$.

\vspace{1.0cm}
We thank the KEKB group for the excellent operation of the
accelerator, the KEK cryogenics group for the efficient
operation of the solenoid, and the KEK computer group and
the National Institute of Informatics for valuable computing
and Super-SINET network support. We acknowledge support from
the Ministry of Education, Culture, Sports, Science, and
Technology of Japan and the Japan Society for the Promotion
of Science; the Australian Research Council and the
Australian Department of Education, Science and Training;
the National Science Foundation of China and the Knowledge Innovation 
Program of Chinese Academy of Sciencies under contract No.~10575109 and IHEP-U-503; 
the Department of Science and Technology of
India; the BK21 program of the Ministry of Education of
Korea, and the CHEP SRC program and Basic Research program 
(grant No. R01-2005-000-10089-0) of the Korea Science and
Engineering Foundation; the Polish State Committee for
Scientific Research under contract No.~2P03B 01324; the
Ministry of Science and Technology of the Russian
Federation; the Slovenian Research Agency;  the Swiss National Science Foundation; 
the National Science Council and
the Ministry of Education of Taiwan; and the U.S.\
Department of Energy.


%

\end{document}